\documentclass[aps,prd,
twocolumn,groupedaddress,
nobibnotes,
nofootinbib,
longbibliography
]{revtex4-1}
\usepackage{amsmath,amsfonts,amssymb,mathrsfs,graphicx,color}
\usepackage{verbatim}
\usepackage{enumerate}
\usepackage{graphicx}
\usepackage{tabularx}
\usepackage[hyperfootnotes=false]{hyperref}
\usepackage{xcolor}
\usepackage{slashed}
\usepackage[utf8]{inputenc}
\usepackage{chngcntr}
\usepackage{siunitx}

\newcommand{\eg}{{\it e.g.}}

\newcommand{\eq}{Eq.}

\newcommand{\fig}{Fig.}

\newcommand{\figs}{Figs.}

\newcommand{\Sec}{Sec.}
 
\newcommand{\App}{Appendix}

\newcommand{\Tab}{Table}

\newcommand{\sibyll}[1]{{Sibyll#1}}
\newcommand{\dpmjet}[1]{{DPMJet#1}}
\newcommand{\qgsjet}{{QGSJet}}
\newcommand{\eposlhc}{{EPOS-LHC}}

\newcommand{\mceq}{MCEq}
\newcommand{\ddm}{DDM}
\newcommand{\rmd}{{\mathrm d}}
\newcommand{\xl}{x_{\rm lab}}

\newcommand{\equ}[1]{\eq~\eqref{eq:#1}}
\newcommand{\figu}[1]{\fig~\ref{fig:#1}}

\newcommand{\bi}{\begin{itemize}}
\newcommand{\ei}{\end{itemize}}

\newcommand{\aeff}{A_\text{eff}}

\newcommand{\revise}[1]{{\color{black}#1}}
\usepackage{commath}

\usepackage{stackengine}
\usepackage{makecell}
\newcommand\brabar{\scalebox{.3}{(}\raisebox{-1.7pt}{--}\scalebox{.3}{)}}

\newcolumntype{Y}{>{\centering\arraybackslash}X}

\graphicspath{{plots/}{}}

\begin{document}

\title{Data-driven hadronic interaction model for atmospheric lepton flux calculations}

\author{Anatoli Fedynitch}
\affiliation{Institute of Physics, Academia Sinica, Taipei City, 11529, Taiwan}
\affiliation{Institute for Cosmic Ray Research, the University of Tokyo,
  5-1-5 Kashiwa-no-ha, Kashiwa, Chiba 277-8582, Japan}
\email{anatoli@gate.sinica.edu.tw}

\author{Matthias Huber}
\affiliation{Technische Universit\"at M\"unchen, Physik-Department, James-Frank-Str.1, D-85748 Garching bei M\"unchen, Germany}

\date{\today}

\begin{abstract}

The leading contribution to the uncertainties of atmospheric neutrino flux calculations arise from the cosmic-ray nucleon flux and the production cross sections of secondary particles in hadron-air interactions. The data-driven model developed in this work parametrizes particle yields from fixed-target accelerator data. The propagation of errors from the accelerator data to the inclusive muon and neutrino flux predictions results in smaller uncertainties than in previous estimates, and the description of atmospheric flux data is good. The model is implemented as part of the \mceq{} package, and hence can be flexibly employed for theoretical flux error estimation at neutrino telescopes.
\end{abstract}

\maketitle

\section{Introduction}
\label{sec:intro}

The interactions of cosmic rays with the Earth's atmosphere create cascades of stable and unstable particles some of which decay into atmospheric leptons \cite{Gaisser:2002jj,Gaisser:2016cr}. From these atmospheric leptons, muons and neutrinos are of particular interest since they serve as a natural ``beam'' for underground large-volume detectors, such as the Super-/Hyper-Kamiokande \cite{Ashie:2005ik,Abe:2018uyc}, the IceCube Observatory with its low-energy extensions DeepCore \cite{Abbasi:2011ym} and IceCube-Upgrade \cite{Ishihara:2019aao}, and the ORCA low-energy array of KM3NeT \cite{Adrian-Martinez:2016fdl}. Above multi-TeV energies, atmospheric neutrinos constitute the main foreground for the characterization of extraterrestrial neutrinos \cite{Aartsen:2016xlq} in IceCube, Antares \cite{Adrian-Martinez:2013bqq}, Baikal-GVD \cite{Avrorin:2011zzc}, and KM3NeT ARCA. For the growing volumes of low-background dark matter experiments and those looking for exotic particles or the diffuse supernova background, atmospheric neutrinos constitute a part of the irreducible background.

Conventional calculations of atmospheric lepton fluxes start from the spectrum and composition of cosmic rays, and track secondary particle cascades down to the ground. The preferred calculation methods are semianalytical solutions of cascade equations \cite{Zatsepin:1962ta,Volkova:1980sw,Gaisser:1983vc,Naumov:1993yr,Lipari:1993hd}, full Monte Carlo calculations (tracking each particle cascade particle individually) \cite{Rancati:1999vb,Barr:2004br,Honda:2006qj,Fedynitch:2012fs}, and iterative numerical solutions \cite{Fedynitch:2015zma} similar to that employed in air-shower simulations (\eg~\cite{Bergmann:2006yz}). At high energies, where the emission angles of neutrinos and muons almost align with the initial cosmic rays, iterative one-dimensional cascade solvers provide high precision and computational speed \cite{Fedynitch:2018cbl}. For low-energy neutrinos below a few GeV, the emission angles of secondary particles in atmospheric cascades play an increasingly important role. These combine with the geomagnetic effects on the cosmic-ray arrival directions and secondary muon trajectories, making the calculation of low-energy neutrino fluxes notoriously challenging. The reference 3D calculations in this energy range \cite{Honda:2015fha,Barr:2004br,Rancati:1999vb} are based on full Monte Carlo simulations that track each secondary particle within the entire volume of the Earth's atmosphere.

While the impact of approximations in the various calculations schemes should be under control \cite{Gaisser:2019xlw}, the theoretical uncertainties of physical models cannot be eliminated. The two dominant uncertainties are the model of hadronic interactions and the parametrization of the cosmic-ray nucleon flux. One approach to characterizing uncertainties is to use inclusive atmospheric muon spectra at energies from GeV scales up to a few TeV to ``calibrate'' particle production yields of hadronic interaction models\footnote{In high-energy physics such models are called event generators} \cite{Honda:2006qj,Honda:2019ymh,Yanez:2019bnw}. An alternative, bottom-up method is the propagation of particle production uncertainties estimated from accelerator measurements through the calculation scheme down to the neutrino fluxes \cite{Barr:2004br,Barr:2006it, Fedynitch_ICRC2017}. Both methods are data driven and thus only produce reliable results within the energy range covered by data. Another basic, model-dependent uncertainty estimation can be obtained by comparing the predictions of multiple models \cite{Fedynitch:2012fs}. In all of these cases, data from particle accelerators or cosmic-ray experiments is not explicitly used in the flux calculation but rather as a reference point for estimating the precision achievable by a hadronic interaction model.

In this work, we develop an empirical data-driven model for the parametrization of secondary particle production, eliminating the impact of phenomenological microscopic models for particle interactions such as Monte Carlo event generators. This method reduces the model dependence in the uncertainty estimation, and produces a data-driven atmospheric lepton flux prediction using a few controllable extrapolations.

\section{Particle interaction models in inclusive flux calculations}

Particle cascades initiated by cosmic rays in the atmosphere of the Earth have been extensively discussed in the literature (see \eg{}, Refs.~\cite{Gaisser:2016cr,Engel:2011zzb,Gaisser:2002jj} for reviews). This work builds upon that of Ref.~\cite{Fedynitch:2018cbl}, which \revise{provides more details on the summary of definitions used below.} For all calculations, we use the public code \mceq{}\footnote{\url{https://github.com/afedynitch/MCEq}}. This section summarizes a few aspects that are relevant for the discussion of hadronic uncertainties in the next sections.

All state-of-the-art flux calculations require some sort of model for secondary particle production in hadronic interactions. For one-dimensional solutions of the transport (cascade) equations [see Eq.~(3) in Ref.~\cite{Fedynitch:2018cbl}], the relevant inputs are the single-differential inclusive production cross sections
\begin{align}
  \begin{split}
    \label{eq:coupling_coeff}
    c_{\ell \to h}(E_\ell,E) &= \frac{1}{\sigma_{\text{inel}, \ell + air}}\frac{\rmd \sigma_{\ell + air \to h + X}}{\rmd E} (E_\ell, E) \\
    &=\frac{\rmd N_{\ell + air \to h + X}}{\rmd E} (E_\ell, E),
  \end{split}
\end{align}
for secondary particles of type $h$ by projectiles of type $\ell$ in collisions with air.

In previous literature, hadronic production yields were discussed in terms of spectrum-weighted moments ($Z$ factors) \cite{Gaisser:2016cr,Lipari:1993hd,Gondolo:1995fq},
\begin{equation}
  \label{eq:z-factor}
  Z_{\text{N}h}(E_\text{N}) = \int_0^1 \rmd{} \xl ~ \xl^{\gamma(E_\text{N}) - 1} \frac{\rmd N_{\text{N} \to h}}{\rmd \xl}(E_\text{N}).
\end{equation}
This energy-dependent scalar function is convenient for semianalytic solutions of cascade equations for the transverse-momentum-integrated (1D) energy spectrum. The longitudinal phase space in $\xl = {E_h}/{E_\text{projectile}}$ is weighted according to the power-law energy spectrum of the projectiles, which for cosmic-ray nucleons is known to fall approximately with $\gamma \approx 2.7$ (for a review, see Chap.~30 in Ref.~\cite{ParticleDataGroup:2020ssz}). Hence, we will often discuss $\xl^{1.7}{\rmd N}/{\rmd \xl}$ for the sake of better visualization of the integrand in Eq.~(\ref{eq:z-factor}) and its clear connection to the relevant phase space.

Most hadronic models in flux calculations are based on tabulated output from event generators or parametrizations of data, and only a few calculations rely on running the full event generators \cite{Rancati:1999vb,Fedynitch:2012fs}. In \mceq{}, \revise{the coefficients $c_{\ell \to h}$ from \equ{coupling_coeff} are calculated by tabulating the output from Monte Carlo event generators}.

The HKKMS models \cite{Sanuki:2006yd} use an inclusive event generator\footnote{``Inclusive event generators'' are programs that simulate the kinematics and multiplicities of secondary particles using probabilities from tabulated inclusive differential cross sections. Single events may violate quantum numbers or energy but \textit{on average} the distributions will converge to the tables. Such models have little in common with the complex conditional probabilities of a full Monte Carlo event generator.} based on tables from \dpmjet{-III} \footnote{\url{https://github.com/DPMJET/DPMJET}} \cite{Roesler:2000he} and the JAM low-energy model \cite{Nara:1999dz}. \textit{Ad hoc} parametrizations inspired by the parton model are introduced to adjust the tables until the muon flux and charge ratio simulations match data to a satisfactory level \cite{Honda:2006qj}. This approach should be sufficiently robust for atmospheric neutrino flux calculations that profit from a sufficient overlap with atmospheric muon and accelerator data. For the high-energy extrapolation, or for observables with weak constraints from muon data, such as \revise{the neutrino-antineutrino ($\nu/\bar{\nu}$) ratio} and the flavor ratio $(\nu_\mu + \bar{\nu_\mu})/(\nu_\text{e} + \bar{\nu_\text{e}})$, a set of physically motivated models might be a more robust choice. \revise{There is ongoing work to improve the interaction model of the HKKMS model \cite{Sato:2021eG}}.

The Bartol calculation \cite{Barr:2004br} uses the inclusive event generator TARGET, which is constructed from phenomenological parametrizations of accelerator data without relying on a microscopic, physical hadronic model. While models like TARGET rely on some empirical assumptions, they can be more precise than \revise{an event generator if the particular phase space is constrained by data, and these data were used to fit the free parameters of the model}. Numerical or analytical calculations can be based on simple table-based models, such as the Kimel-Mokhov model \cite{Kimel:1974sn} which, despite its age, produces meaningful high-energy fluxes \cite{Kochanov:2008pt,Sinegovskaya:2014pia}.

\section{Particle production phase space}

\label{sec:phase_space}
\begin{figure*}
  \centering
  \includegraphics[width=.96\textwidth]{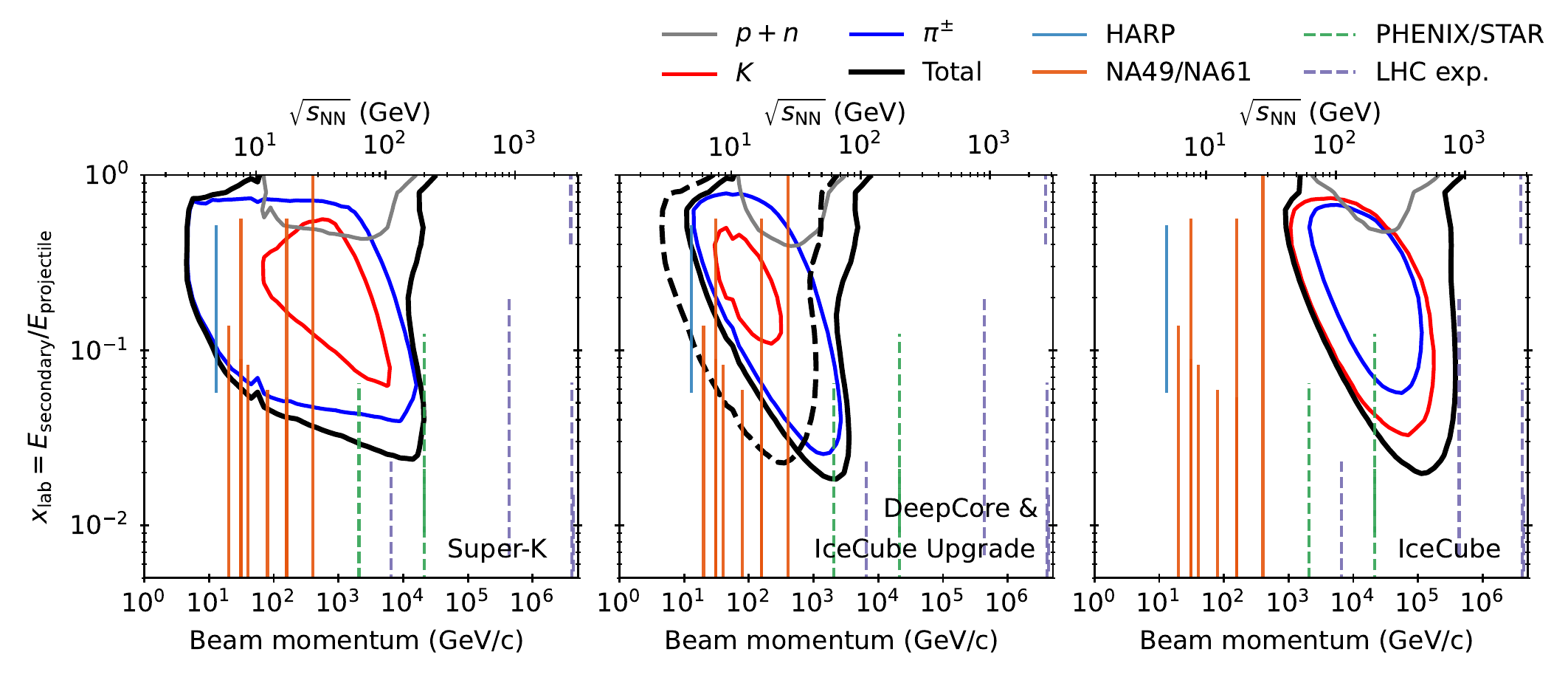}
  \includegraphics[width=.96\textwidth]{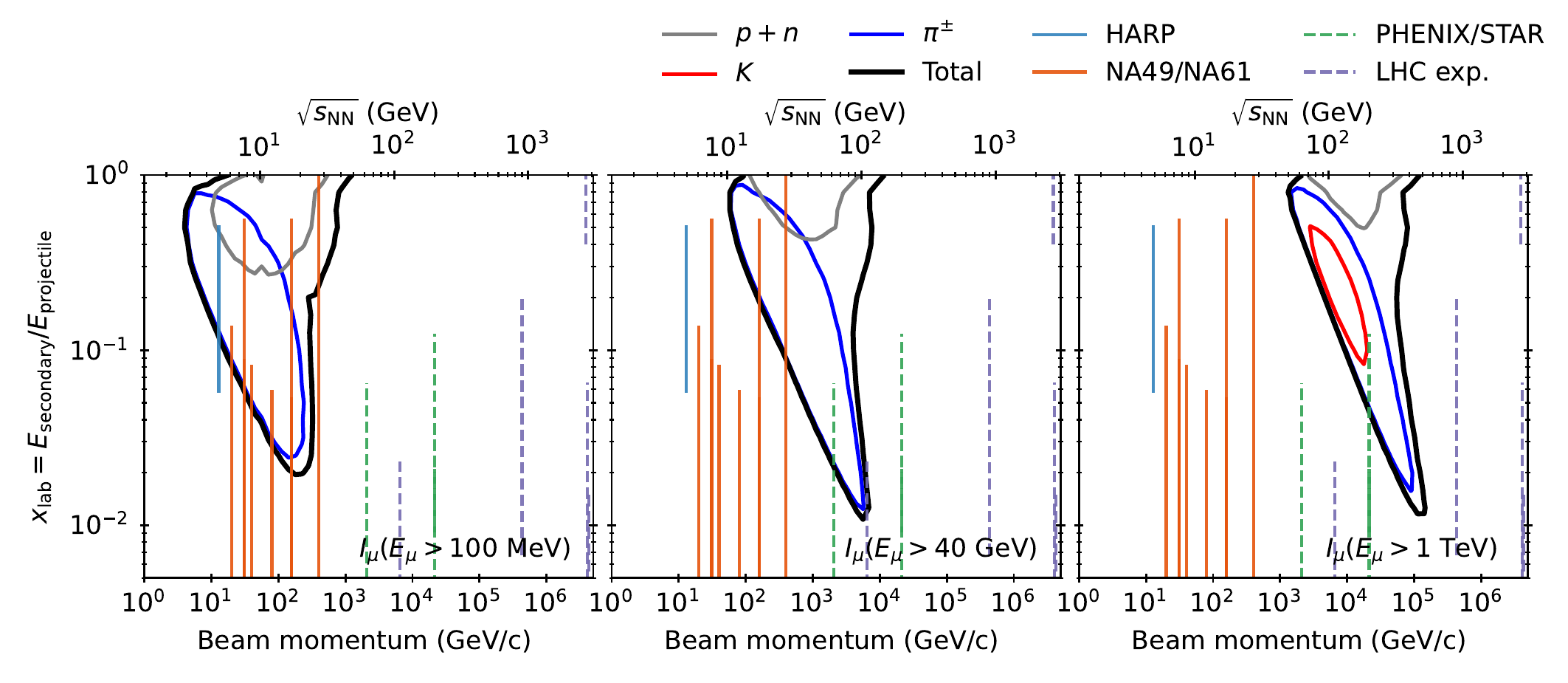}
  \caption{The top three panels show the particle production phase space sampled by atmospheric neutrino experiments. The event-rate computations are explained in Appendix \ref{app:effective_areas}. The black contour represents the phase space responsible for 90\% of the event rate. The colored contours are fractions of the rate originating from specific secondary particles. The total phase space probed by an experiment, or muon intensity level is shown by thick black curves. Similar contours as for IceCube can be expected for the high-energy neutrino observatories Baikal GVD, KM3NeT ARCA, and P-ONE. The DeepCore (solid) and IceCube-Upgrade (dashed) contours have been obtained for a reconstructed energy $E_\text{reco} < 60$ GeV, a typical value for atmospheric oscillation analyses. The bottom panels show the particle production phase space accessible through inclusive atmospheric muon measurements. The contours enclose regions contributing to the rate of surface muons above an indicated threshold energy. Differential spectrometer data can cover a wider range of contours as shown for the integral fluxes. The vertical lines have been adopted from Fig.~18 of Ref.~\cite{Albrecht:2021cxw} and indicate the kinematical acceptance for $pp \text{or} pA \to \pi^\pm + X$ at fixed target experiments (solid vertical lines) and colliders (dashed). The visible LHC experiments are those with forward coverage and particle identification capability, namely SMOG@LHCb ($\sqrt{s} =200$ GeV), CMS, ATLAS, LHCb, and LHCf.}
  \label{fig:phase_space_neutrinos_muons}
\end{figure*}
\begin{figure}
  \centering
  \includegraphics[width=1\columnwidth]{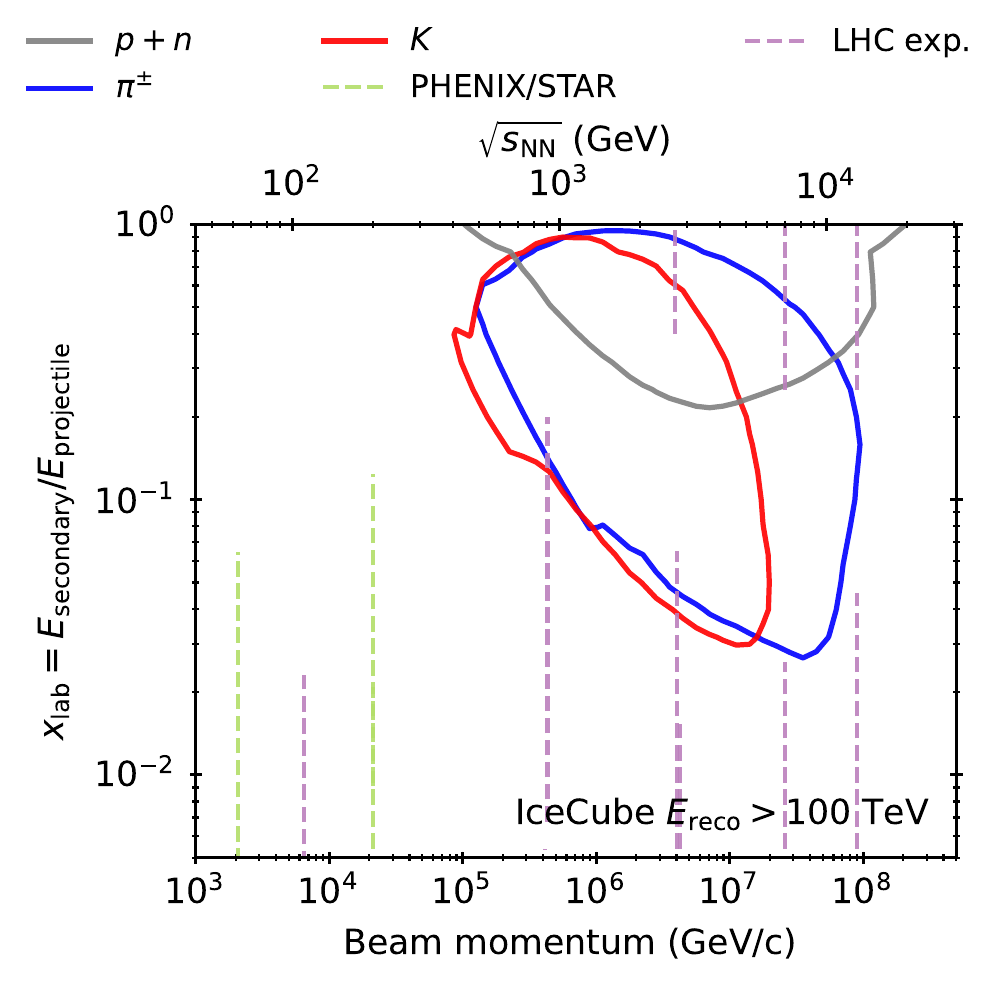}
  \caption{Same as the rightmost panel of \fig~\ref{fig:phase_space_neutrinos_muons} but for events with a reconstructed energy $E_\text{reco} > 100$ TeV, which is a more relevant energy range for astrophysical flux and prompt neutrino analyses.}
  \label{fig:phase_space_conv_100TeV}
\end{figure}

\begin{figure}
  \centering
  \includegraphics[width=1\columnwidth]{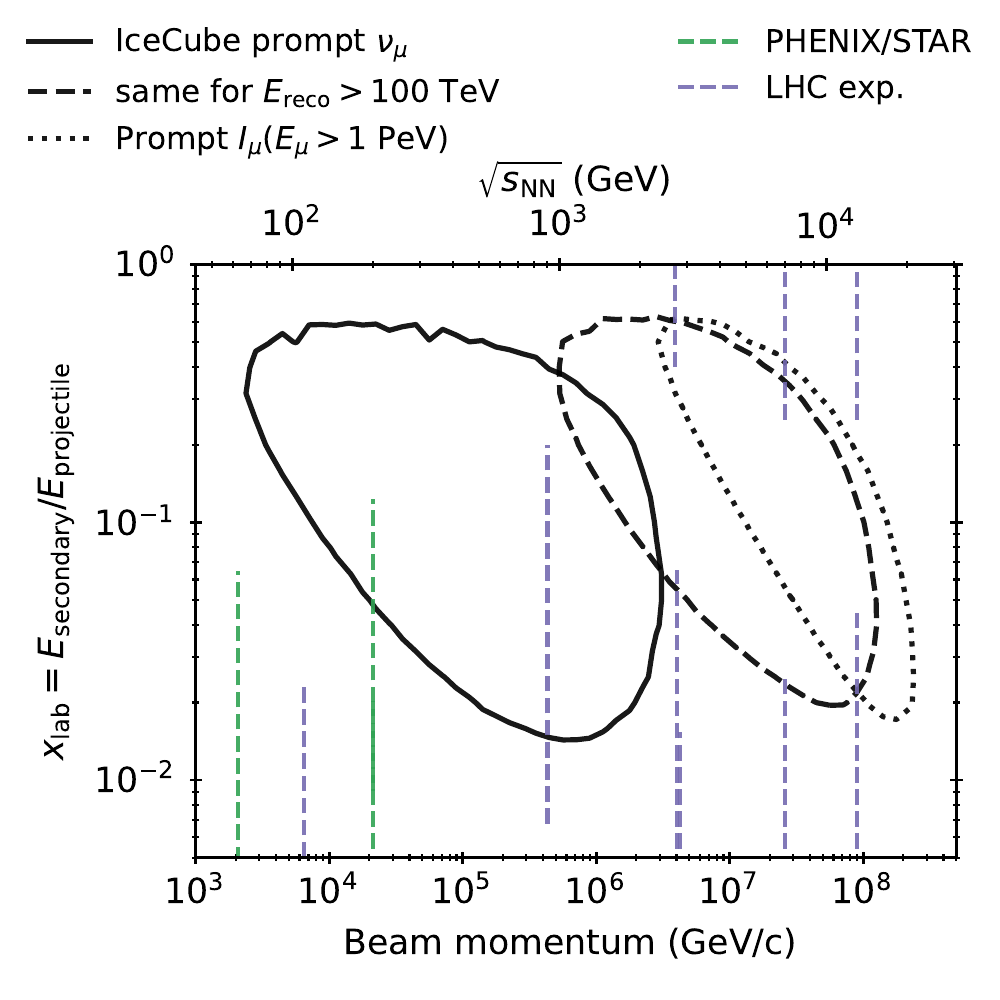}
  \caption{Charged and neutral $D$-meson phase space contributing to 90\% of all prompt neutrino tracks in IceCube (solid), and tracks with reconstructed energy above 100 TeV as an example for a more realistic search window (dashed). The dotted contour shows a hypothetical prompt muon-rate measurement, half of which originates from unflavored mesons decays and the remaining \revise{half} from charm. The detector acceptance (illustrated as vertical lines) has been calculated assuming pion secondaries as in \fig~\ref{fig:phase_space_neutrinos_muons}.}
  \label{fig:phase_space_prompt}
\end{figure}

As discussed in Sec.~IV.B of Ref.~\cite{Fedynitch:2018cbl} (or, \eg{},~in \cite{Gaisser:2016cr}), the hadrons with the highest relevance for atmospheric lepton production are those with a high production yield and high branching ratios into leptons. The phase space for atmospheric neutrino production has been studied in detail by several authors \cite{Engel:1999zq,Barr:2006it,Sanuki:2006yd}, and most recently in Ref.~\cite{Honda:2019ymh} for conventional leptons and in Ref.~\cite{Jeong:2021vqp} for prompt neutrinos. For conventional fluxes, these are charged pions, and charged and neutral kaons. Prompt neutrino fluxes originate from charmed or bottom mesons. Although nucleons do not decay into leptons directly, very forward nucleon yields ($\xl \gtrsim 0.6$) affect inclusive lepton fluxes due to modifications to the regeneration $Z$ factors, $Z_{NN}$ (where $N$ for proton or neutron), that can shift the average production altitude and modify the contribution from secondary particle interactions. The nucleon yield and the inelasticity have a higher impact on (exclusive) air showers, where interactions of low-energy particles, strange baryons, and antibaryons at lower altitudes play a more important role \cite{Pierog:2006qv}.

Figure~\ref{fig:phase_space_neutrinos_muons} shows the two-dimensional phase space that gives rise to 90\% of the events in atmospheric neutrino detectors. The muon neutrino rates in Super-K (cf.~Fig.~1 in Ref.~\cite{Barr:2006it}), Hyper-K, and the IceCube-Upgrade receive contributions from particle interactions just above the inelastic threshold. The contours of low-energy atmospheric neutrino detectors have sufficient overlap with the phase space covered by fixed-target detectors. The yields of kaons provide significant contributions at equivalent beam energies above 80 GeV, well in reach of the NA61 experiment. For \textit{conventional} atmospheric events in cubic-kilometer scale detectors (rightmost panel) there is almost no accelerator data, in particular from forward detectors with particle identification capabilities. The phase space probed by IceCube lies within approximately $\sqrt{s} \sim\SIrange[range-phrase=-, range-units=single]{100}{900}{GeV}$, an energy range corresponding to that of RHIC and the Sp$\bar{p}$S. This is significantly lower than the modern LHC beam configurations. The energy range has been extensively studied at the CERN S$p\bar{p}$S accelerator, and it might be worth to investigate the possibility of using these data in a later study. With higher cuts on $E_\text{reco}$ in IceCube, the contours are pushed to higher energies within the range of LHC beam energies (see \fig~\ref{fig:phase_space_conv_100TeV}). However, most of this phase space is either not instrumented or lacks charged particle identification capabilities. In the future, the FASER experiment \cite{Kling:2021gos} or the proposed Forward Physics Facility (FPF) \cite{Feng:2022inv} will attempt to provide direct constraints on forward neutrino fluxes.

The still unobserved \textit{prompt} neutrino rate in IceCube probes collisions up to a center of mass energy of $\sim 1$ TeV, shown as a solid contour in \fig~\ref{fig:phase_space_prompt}. Given sufficient luminosity and a small uncertainty, a $D$-meson spectrum measurement by LHCb at 900 GeV, 2.76 GeV, and 7 TeV for proton-oxygen collisions would cover a sufficient cross section to constrain some of the large prompt flux uncertainties. To obtain tight experimental constraints on conventional and prompt atmospheric fluxes, the LHC has to be operated at the lowest possible energies during the foreseen proton-oxygen collision runs with the proton fragmentation zone pointing toward LHCb.

\begin{figure}
  \centering
  \includegraphics[width=\columnwidth]{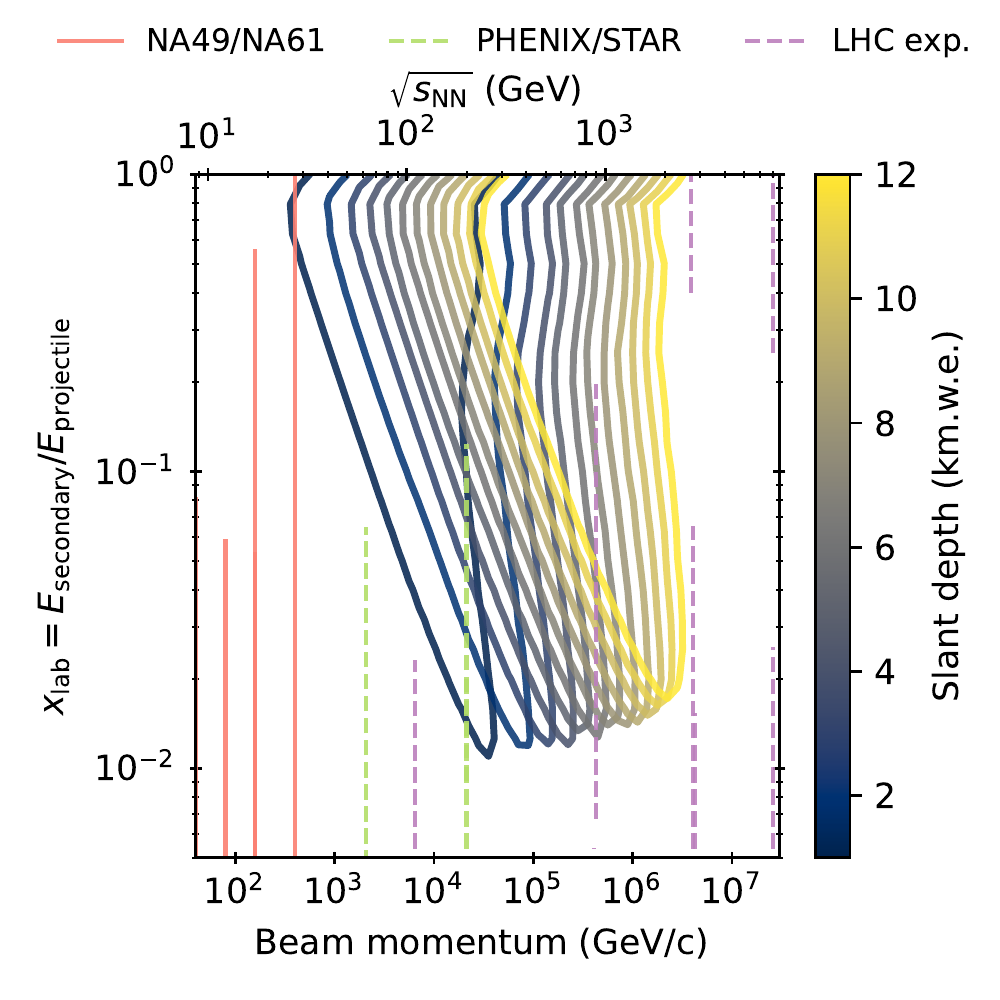}
  \caption{Phase-space coverage of deep underground muon rate measurements. The contours cover 1--12 kilometer-water-equivalent slant depth in standard rock. Note that the black IceCube contour in \fig~\ref{fig:phase_space_neutrinos_muons} entirely overlaps with the phase space accessible at underground laboratories. The very-high-energy selection shown in \figs~\ref{fig:phase_space_conv_100TeV} and \ref{fig:phase_space_prompt} are only partially covered by underground muon measurements.
  \label{fig:phase_space_underground}}
\end{figure}
In the absence of data from proton-oxygen collisions, tighter constraints on forward light meson production can be obtained from atmospheric muons, which have been measured with precision a comparable to or exceeding that of forward detectors at accelerators. The bottom panels in \fig~\ref{fig:phase_space_neutrinos_muons} show the corresponding contours for the rate of muons at the surface above several energy thresholds. In the Super-K and DeepCore energy range, atmospheric muons constrain muon neutrino fluxes almost model independently since both originate from pion decays and share the same initial cosmic-ray spectrum. To cover the IceCube energy range, the required muon energy at the surface is $>1$ TeV. More \revise{relevant} are muon fluxes or rates observed in deep underground detectors, which are known to much better precision than the few-TeV-range measurements at the surface \cite{Fedynitch:2021ima}. While there is full overlap between the deep underground contours in \fig~\ref{fig:phase_space_underground}, the caveat is that only $20$--$30\%$ of muons observed at large depths originate from kaon decays (see Fig.~5 in Ref.~\cite{Fedynitch:2021ima}), in contrast to almost 80\% of observed muon neutrinos. Under some soft model dependence, data-driven constraints from underground muons on very-high-energy conventional neutrino fluxes will have much lower uncertainties than estimates (such as Ref.~\cite{Barr:2006it}) that imply the complete absence of data from accelerators.

For nucleons (thin gray curves in \fig~\ref{fig:phase_space_neutrinos_muons}), the relevant phase space is very forward ($\xl \gg 0.5$), \revise{, contributing significantly} to the inelasticity and $Z_{NN}$. Lower inelasticities deepen atmospheric cascades, resulting in less energy dissipation into high-$\xl$ secondaries during the first few cascade generations. The impact of baryon yields on the \revise{energy} spectra is small and almost featureless. Thus, any nondegenerate constraints on forward baryon production are unlikely to be obtained from atmospheric leptons alone.

\section{Data-driven Hadronic Interaction Model (\ddm{})}
\label{sec:\ddm{}}

This section discusses available fixed-target data and reviews the requirements for inclusive hadronic interaction models.

An inclusive hadronic interaction model is a set of tables or parametrizations of differential secondary particle yields and interaction cross sections with the following requirements:
\begin{enumerate}
  \item Wide projectile interaction energy range (see \Sec~\ref{sec:phase_space}):
        \begin{enumerate}[(a)]
          \item From particle production threshold up to a few hundred TeV for multi-kton to Mton atmospheric neutrino detectors, such as DeepCore/IC Upgrade, KM3NeT ORCA, Super-K, Hyper-K, DUNE, etc.
          \item 100 GeV to 100 PeV or higher for IceCube (Gen-2 \cite{Aartsen:2020fgd}), KM3NeT ARCA, Baikal GVD \cite{Avrorin:2013uyc}, and P-ONE \cite{Agostini:2020aar}.
        \end{enumerate}
  \item Supports $p$, $n$, $\pi^\pm$, $K^0_{L|S}$, and $K^\pm$ as projectiles and provide inclusive production cross sections for the same particles.
  \item Target nuclei are close to the average mass number of air, $A \approx 14.5$. For inclusive fluxes, the difference between carbon and nitrogen targets is less than 2\%.
  \item The secondary particle yields are differential in
        \begin{enumerate}[(a)]
          \item $\xl$ for one-dimensional, or
          \item $p_z$ and $p_\perp$ (or the scattering angle $\theta$) for three-dimensional calculations.
        \end{enumerate}
  \item Errors and covariance matrices \revise{for the free parameters or the data}.
\end{enumerate}

In the following, we develop a new one-dimensional model differential in $\xl$ for nucleon and pion projectiles that aims to address most points using published data from accelerators.

\subsection{Data selection}
\label{sec:data_selection}
\begin{figure*}
  \centering
  \includegraphics[width=\textwidth]{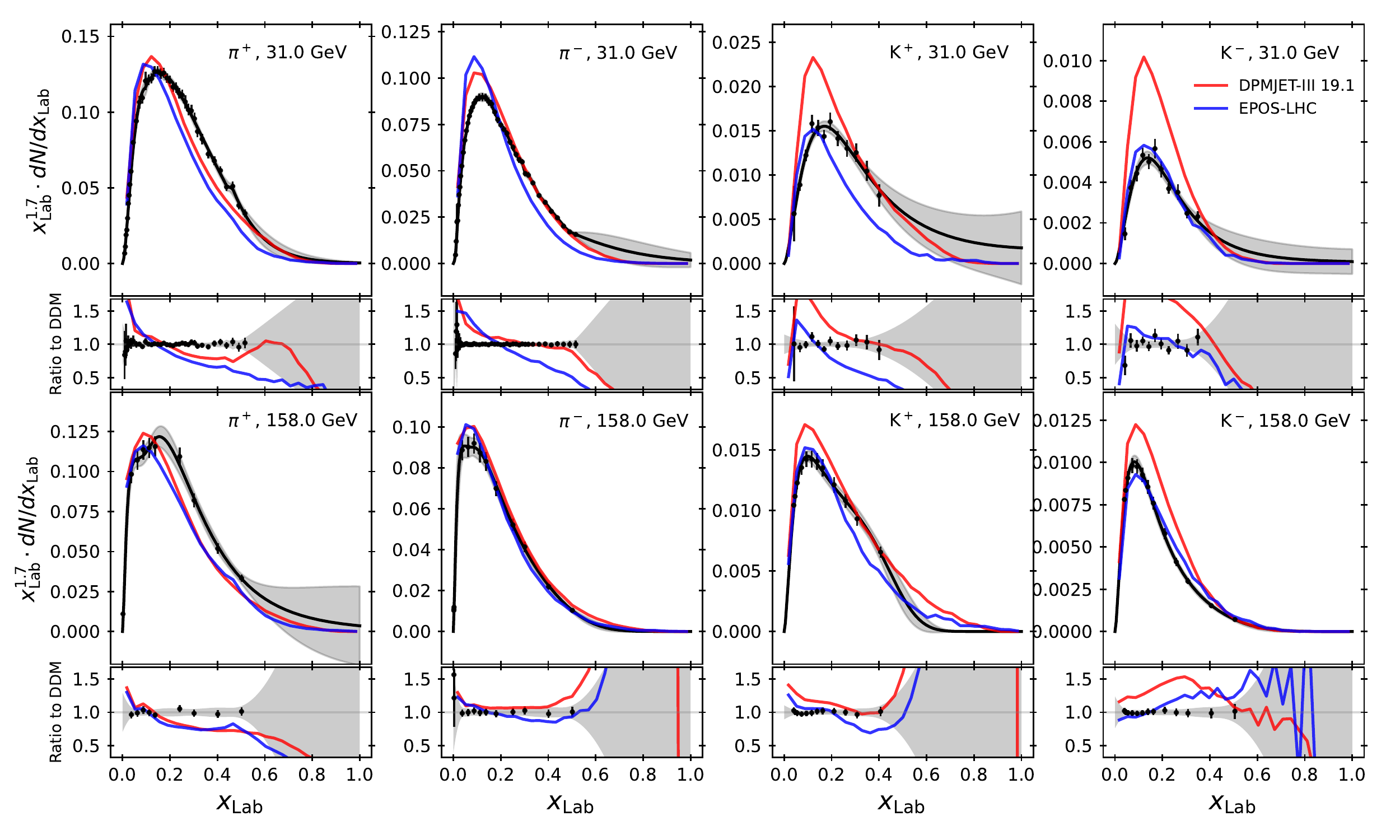}
  \caption{Parametrization of light meson yields in the \ddm{}. The data points in the upper panels are for proton-carbon collisions at 31 GeV from NA61/SHINE \cite{Abgrall:2015hmv}, integrated over scattering angle. The lower panels show NA49 proton-carbon data at 158 GeV, transformed into the laboratory frame and integrated over $p_\perp$ \cite{Alt:2006fr}. The data points for K$^\pm$ are extrapolated to proton-carbon from NA49 proton-proton data \cite{Anticic:2010yg} (see Appendix \ref{app:pp_pc_extrapolation}). Reference hadronic interaction models are shown in color and differences appear larger due to the linear scale and the factor $\xl^{1.7}$ that emphasizes the relevant phase space for the $Z$ factor integrals in \equ{z-factor}.}
  \label{fig:pC_data_mesons}
\end{figure*}

\begin{figure}
  \centering
  \includegraphics[width=.9\columnwidth]{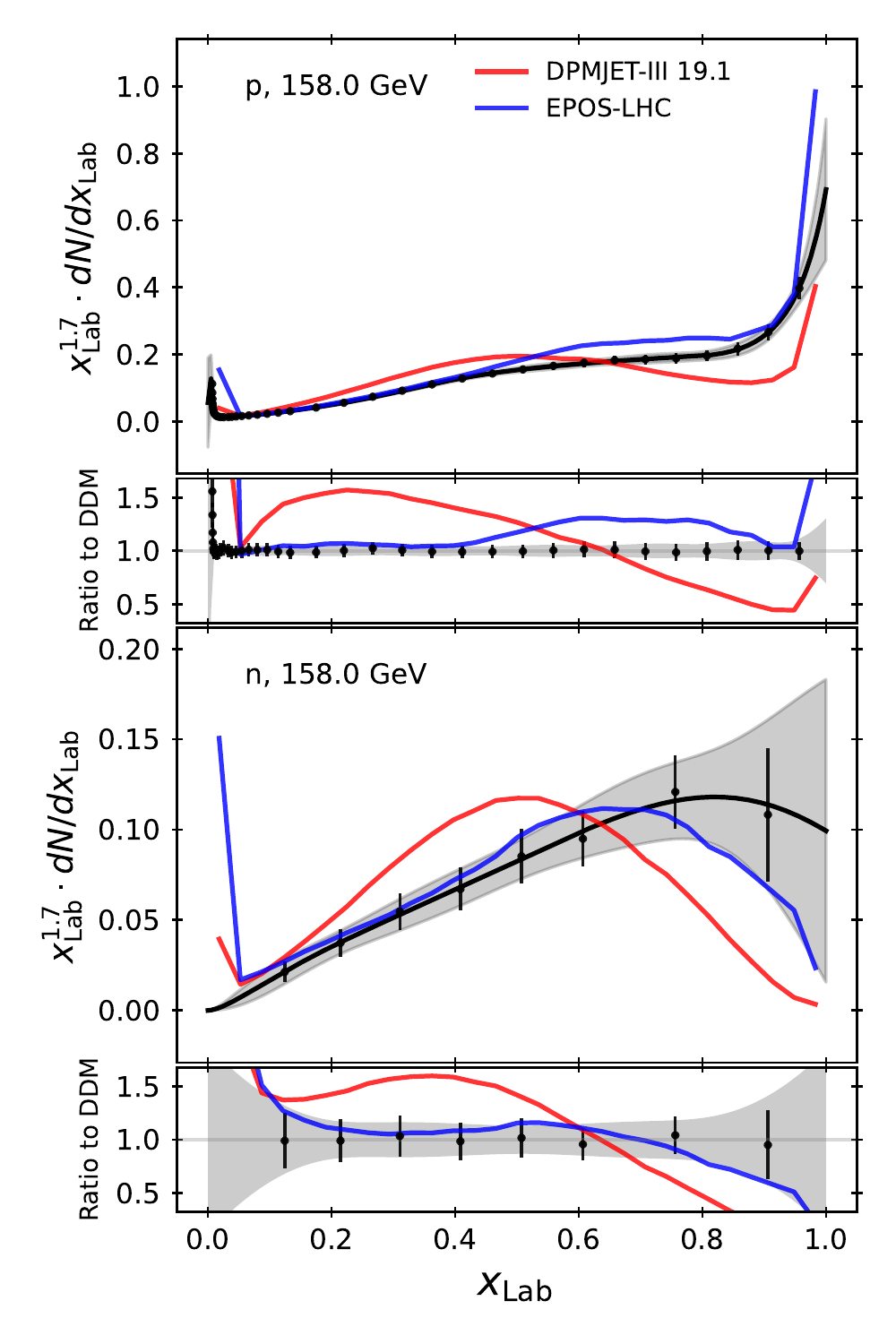}
  \caption{Parametrization of proton and neutron spectra based on NA49 proton-carbon data at 158 GeV \cite{Alt:2006fr}. \dpmjet{} shows a clear deficit of baryons at $\xl\sim0.8$, similar to what has been found in \sibyll{-2.1} and improved in \sibyll{-2.3d} \cite{Riehn:2019jet}. The baryon distributions are crucial for the definition of inelasticity and have greater impact on the distribution of shower maxima.}
  \label{fig:pC_data_baryons}
\end{figure}

\begin{table}
  \centering
  \begin{tabular}{lcccc}
    \hline
    Experiment & beam                  & $E_{\textrm{beam}}$/GeV & Secondaries                                              & Variables             \\
    \hline
    NA49       & $p$C                    & 158                     & $\pi^{\pm}, \stackon[.3pt]{p}{\brabar},
    n$         & $x_\text{F}, p_\perp$                                                                                                              \\
    NA49       & $pp$                    & 158                     & K$^{\pm}$                                                & $x_\text{F}, p_\perp$ \\
    NA61/SHINE & $p$C                    & 31                      & $\pi^{\pm}, \text{K}^{\pm}, \text{K}_\text{S}^0, \Lambda
    $          & $p, \theta$                                                                                                                        \\
    NA61/SHINE & $\pi^-$C              & 158, 350                & $\pi^{\pm}, K^{\pm}, \stackon[.3pt]{p}{\brabar}
    $          & $p, p_\perp$                                                                                                                       \\
    \hline 
  \end{tabular}
  \caption{Summary of particle yield measurements from NA49 \cite{Alt:2006fr,Anticic:2010yg} and NA61/SHINE \cite{Abgrall:2015hmv,Prado:2018wsv}. The \ddm{} is built using the double-differential yields in the variables indicated in the last column.}
  \label{tab:na_data}
\end{table}

As discussed in \Sec~\ref{sec:phase_space}, a part of the particle production phase space is probed by accelerator experiments in fixed-target configurations. A screening of data was previously performed for the Bartol neutrino flux calculation \cite{Barr:2004br,Barr:2006it}, which focused on Super-K range of energies ($E_\nu < 1\, \mathrm{TeV}$). The most notable data releases since then are the final results of NA49 and the still incoming results of its successor NA61/SHINE \cite{Abgrall:2014xwa}. The relevant runs for the present work are those performed  with thin carbon targets that lie close to the average mass number of air. The analyzed and published single- and double-differential cross sections from these experiments are listed in \Tab~\ref{tab:na_data}. Data taken with limited forward acceptance, or binned in rapidity\footnote{The energy ramp in proton-proton interactions \cite{Aduszkiewicz:2017sei} could have been very useful to study the energy dependence of particle yields and the onset of Feynman scaling for each particle species. However, the errors on the differential $\xl$ spectrum after integration in the $(y, p_\perp)$ plane are too large due to the large measurement uncertainties and the limited detector acceptance of this run. Thus, no significant scaling trend has been identified within the errors.} instead of momentum, has a limited impact on this study since there is insufficient acceptance at high $x_\text{F}$ at low $p_T$ values. NA61/SHINE has taken more data between 30 and 160 GeV in proton-carbon and pion-carbon interactions but the results have not yet been published at the time of writing.

The Data Driven Model (\ddm{}) is exclusively based on the sets in \Tab~\ref{tab:na_data}, taken with thin carbon targets. As pointed out in Ref.~\cite{Fedynitch:2018cbl}, the absence of a charged kaon analysis for proton-carbon at 158 GeV in NA49 and NA61/SHINE is essential and requires a workaround. We use charged kaon data from $pp$ collisions at NA49 \cite{Anticic:2010yg} and extrapolate the data to proton-carbon using a combination of interaction models (see \App~\ref{app:pp_pc_extrapolation}). Data from NA56/SPY \cite{Ambrosini:1999id} may further help to constrain the model at high energies but it requires a more complex assessment of uncertainties related to the extrapolation from beryllium to a carbon or air target, the medium thickness of the target, and the limited angular detector acceptance.

Data from colliders could be helpful to assess the high-energy extrapolation uncertainties of the \ddm{}. But the limitations on forward acceptance and larger errors of forward detectors only marginally probe the $x_\text{F} > 0.1$ phase space at $\sqrt{s} > 1$ TeV. Older measurements from \revise{Intersecting Storage Rings or S$p\bar{p}$S} suffer from additional uncorrected errors, such as feed-down from strange baryons \cite{Anticic:2010yg}, although recently these corrections have been performed for some older proton-proton data sets \cite{Fischer:2022oqp}. Indirect constraints can come from the zero-degree calorimeter experiments LHCf and RHICf \cite{Adriani:2008zz}, which measure neutral particles within a narrow $p_\perp$ range at $x_\text{F} = \SIrange[range-phrase=-, range-units=single]{0.5}{1}{}$. One important result is the confirmation of Feynman scaling \cite{Feynman:1969ej} at LHC energies \cite{Adriani:2015iwv} for small $p_\perp$. As previously discussed, good collider constraints would come from air-shower specific measurements at LHCb in proton-oxygen runs \cite{Albrecht:2021cxw}. An alternative source of constrains are atmospheric inclusive muons \cite{Sanuki:2006yd,Yanez:2019bnw}, deep underground muons \cite{Barrett:1952woo,Mei:2005gm,Bugaev:1998bi,Fedynitch:2021ima}, seasonal variations \cite{Adamson:2007ww,Heix:2019jib} and atmospheric neutrinos \cite{Fedynitch:2019bbp}.

\subsection{Parametrization of data and its uncertainties}
\label{sec:uncertainty_parametrization}

The NA49 data is provided in the center-of-mass-frame variable $x_\text{F} \approx p_z/\sqrt{s}$ and requires a transformation into the target's rest frame. This is done by fitting the $p_\perp$ distribution in each $x_\text{F}$ bin using
\begin{equation}
  \frac{\rmd n}{\rmd p_\perp} = a_0 p_\perp^{a_1}~e^{a_2p_\perp^{a_3}},
\end{equation}
and a bootstrap method to convert from the ($x_\text{F},p_\perp)$ to $(\xl, p_\perp)$. The single-differential $\xl$ distribution is obtained by integrating over $p_\perp$. The same method is used to propagate the experimental errors, approximated as the geometrical sum of the statistical and systematic errors. The NA61 data set is published as a function of scattering angle and total laboratory momentum $(\theta, p)$, and hence single-differential distributions can be readily obtained through integration over $\theta$.

\figs~\ref{fig:pC_data_mesons} and \ref{fig:pC_data_baryons} show the meson and baryon yields, respectively. Similar fits for $\pi^-$-carbon data have been obtained from NA61, and some problems are discussed in Appendix \ref{app:pi-c-fits}. The natural logarithm of the data is fit using cubic splines\footnote{\textit{splrep} function from SciPy \cite{2020SciPy-NMeth}}, except for $\pi^\pm$ at 31 GeV, which requires linear splines for robust fits. A smoothing factor $s>0$ is chosen such that the fit follows all trends in the data, and the error on the $Z$ factor does not significantly change for larger values of $s$. The best fit, and thus the central value of the predicted atmospheric lepton fluxes, is sufficiently robust against changes to $s$ and the choice of the spline order since the phase space contributing most of the $Z$ factor is well covered by data. For the computation of \mceq{} interaction matrix coefficients in \equ{coupling_coeff}, the \ddm{} splines are numerically averaged within each logarithmic energy bin.

The spline uncertainties are derived from the covariance matrix, which is obtained from the Hessian matrix computed using finite differences. The \revise{chosen value of $s$} defines the number of knots and influences to which extent features in the data smear out and what the size of the resulting error band is. To improve the containment of the experimental $1\sigma$ error bars within the uncertainty bands of the splines, the covariance matrix has been multiplied by factor 2. \revise{When using splines, some empirical choices have to made to avoid the case $s\to0$, for which the splines turn into interpolating splines with zero errors at the data points}. By comparing the ratio panels in Figs.~\ref{fig:pC_data_mesons} and \ref{fig:pC_data_baryons}, it can be seen that the errors increase swiftly in the absence of data where the model extrapolates. Therefore, the total uncertainty, especially that of the Z-factor, depends quite significantly on the position of the rightmost data point. \revise{We investigated that one additional data point at higher $\xl$ for $\pi^+$ and $K^+$ at 158 GeV significantly reduces the extrapolation uncertainty, even if one assumes a larger error}.

A more rigorous or robust approach has not been found due to the conceptual problem of fitting and extrapolating data in the absence of a physical model. Forward particle yields probe the nonperturbative regime, which is not consistently well described by the hadronic models (see colored curves). The differences between the two models are larger than the experimental uncertainty and that of the splines. Thus, uncertainty estimates based on ``bracketing'' different models should in most cases result in an overestimation of errors. Instead of splines, we attempt fits with empirical functions similar to those used in the TARGET model of the Bartol calculation \cite{Engel:1999zq,Robbins:2004sy,Barr:2004br}. Due to the imposed shape of the function and fewer parameters, the extrapolation to large $\xl$ is overconstrained, resulting in too small errors given that the particle yields at very large $\xl$ are experimentally not known. Since the aim of the \ddm{} is to parameterize the data \textit{and} its uncertainties, empirical functions are discarded due to the imposed bias. On the other hand, splines can only be applied where sufficient data is available. As \revise{discussed in \App~\ref{app:pi-c-fits} and shown} in Table~\ref{tab:$z$-factors}, the present spline fit method struggles to describe $K^\pm$ production in pion-carbon data due to the limited experimental phase space. In this case, a best fit can be easily found but the $Z$ factor integral errors do not converge. As shown by Fischer \textit{et al.}~in Ref.~\cite{Fischer:2022oqp} for sufficiently abundant proton-proton data, splines can be used for cross calibrating experiments that individually cover small patches of phase space to obtain ``global spline fits'' similar to that of the Global Spline Fit (GSF) \cite{Dembinski:2017zsh} for cosmic-ray fluxes.

\subsection{Model assumptions}

\begin{figure*}
  \centering
  \includegraphics[width=0.95\textwidth]{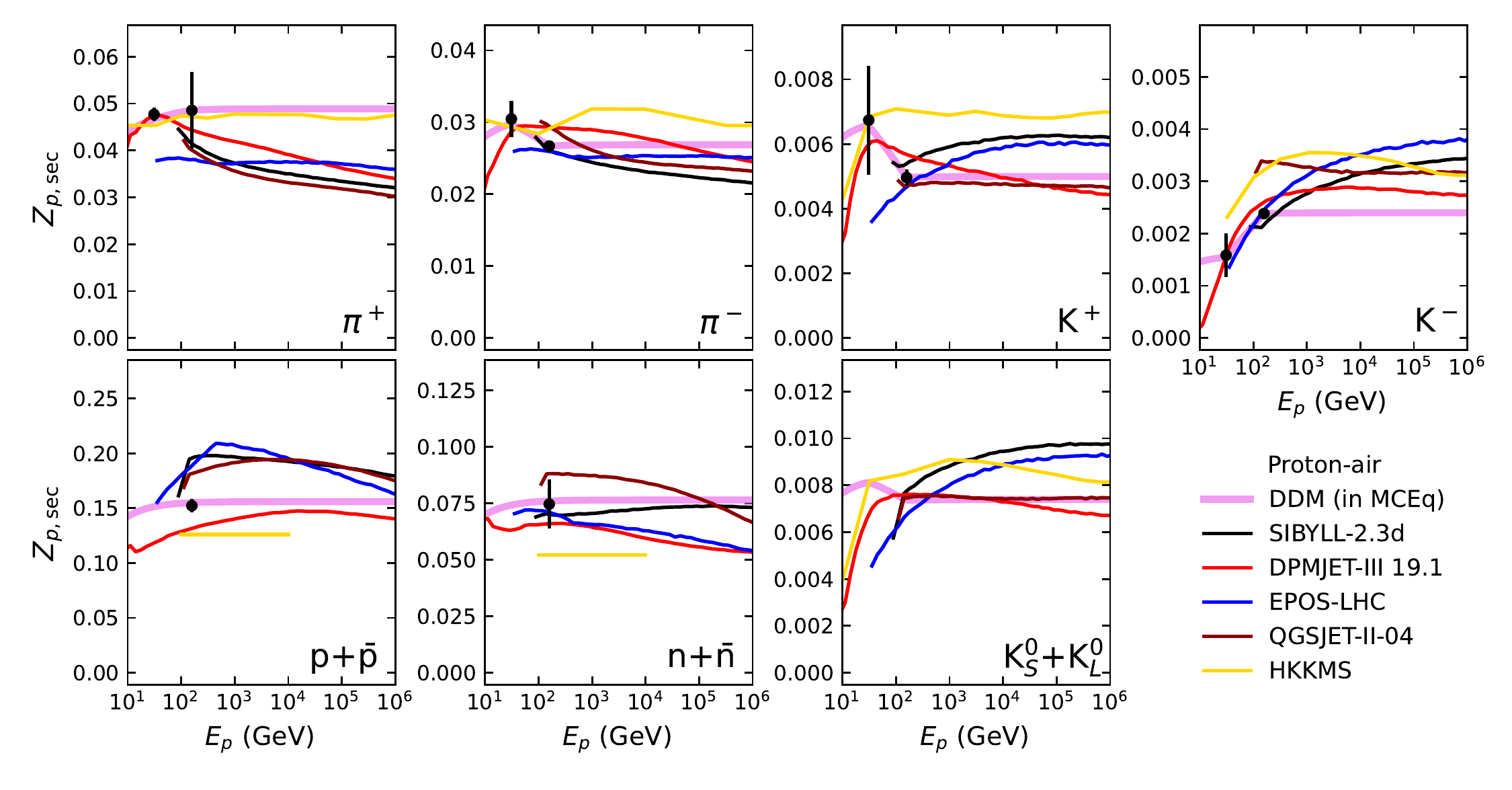}
  \caption{Energy-dependent spectrum-weighted moments ($Z$-factors) computed for an air target and $\gamma = 1.7$ according to \equ{z-factor}. Tabulated values for other $\gamma$ are located in Appendix \ref{app:z-factor-table}. The result of the HKKMS interaction model, which is based on an older version of \dpmjet{-III} and has been tuned to inclusive muons \cite{Sanuki:2006yd}, is shown in gold. Strict Feynman scaling, recognizable as approximate energy independence, is not favored by most of the models. In the \ddm{}, scaling appears by construction above 158 GeV (see text). In HKKMS, scaling is a result of tuning. Although \dpmjet{} demonstrates good compatibility with the shown data points, forward yields at higher energies consistently decline and are chosen to be corrected in HKKMS. A similar decline is visible in data for $\pi^-$ and $K^+$, while for $\pi^+$ a conclusion cannot be drawn due to the error on data. K$^0$ are derived from isospin relations and line up well with the \dpmjet{} and \qgsjet{} models. The differences between models are consistently much larger than the error on data.}
  \label{fig:zfactors}
\end{figure*}
\begin{figure*}
  \centering
  \includegraphics[width=0.9\textwidth]{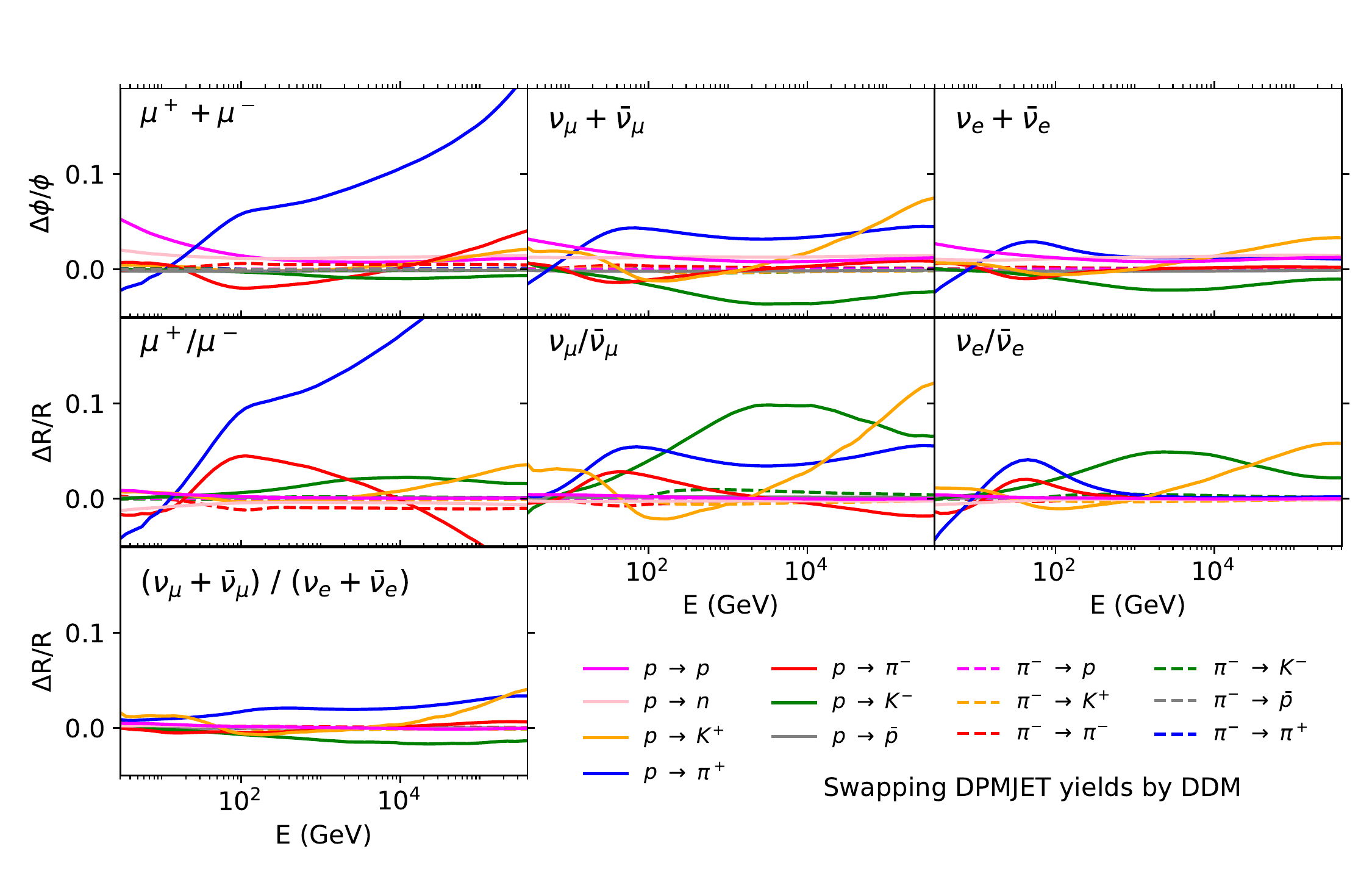}
  \caption{Effect on inclusive fluxes from each channel in the \ddm{}. Here, a standard calculation with \mceq{} using the \dpmjet{-III 19.1} model is modified by swapping individual hadronic channels with those from the \ddm{}. The dominant impact comes from the $\pi^+$ yields at higher energies (blue solid curves). The dashed curves demonstrate that secondary pion interactions, even if substantially modified (see Appendix \ref{app:pi-c-fits}), are not very relevant for inclusive flux calculations.}
  \label{fig:impact_channels}
\end{figure*}

A consistent inclusive interaction model is constructed starting from an initial library of particle yields from \dpmjet{}. Particle yields known to the \ddm{} are replaced, while the remaining very rare production channels are retained from \dpmjet{}. We verified that the results only marginally ($\sim1\%$) change for initial model choices other than \dpmjet{}. Figure~\ref{fig:zfactors} contains the energy-dependent spectrum-weighted moments computed from the data in Tab.~\ref{tab:na_data} and various current hadronic interaction models. The $Z$ factors are a sufficient framework to discuss extrapolation uncertainties.

The strongest assumption in the \ddm{} is Feynman scaling (FS) \cite{Feynman:1969ej}. In simplified terms, the idea is that once partons scatter and form color chains (or strings), there is a universal minimal cost to pull new partons from the vacuum if the critical string tension is exceeded. At higher collision energies, the longitudinal phase space grows but the number of secondaries per phase-space element is constant. As a consequence, the longitudinal momentum spectrum in the scaling variable $x_\mathrm{F}$ is independent of energy. Although this may be a very simplified description of the complexity of hadron scattering, the idea catches the essentials of the nonperturbative modeling of interactions. FS is approximately realized in data and it is in particular well motivated at energies where multiple partonic interactions have little effect. Within a limited $(\eta,p_\perp)$ range, LHCf demonstrated that FS also holds at LHC energies \cite{Adriani:2015iwv}. FS is known to be violated due to the significant contribution of hard processes at central rapidities and high energies due to multiple partonic interactions. Some violation of forward scaling is also expected due to, e.g., the energy dependence of diffractive cross sections and significant contributions of resonances to the inclusive yields of light hadrons \cite{Fedynitch:2018cbl,Riehn:2019jet,Anticic:2010yg}. The \ddm{} $Z$ factors for negative pions in Fig.~\ref{fig:zfactors} indicate a violation of scaling between the 31 and 158 GeV data (black circles). For kaons, this cannot be stated with sufficient significance since the 31 GeV beam energy is too close to the production threshold.

Nonetheless, we assume FS for the \ddm{} above 158 GeV for three reasons: 1) the FS violation in central or hard scatterings is suppressed for inclusive fluxes due to the factor $\xl^{\gamma-1}$ in \equ{z-factor}; 2) there is no clear, consistent trend in data and in the event generators; 3) assuming an additional {\it ad hoc} error is another source of bias in the absence of a physical model, as discussed in \Sec~\ref{sec:uncertainty_parametrization}. In Ref.~\cite{Barr:2006it} a (pessimistic) {\it ad hoc} extrapolation error was assumed. Since only data at two beam energies are available, the \ddm{} interpolates between the 31 GeV and 158 GeV data linearly in $\log(E_\textrm{p})$. Once new data are released by NA61, these can be included for a more sophisticated transition and serve as an additional cross-check. At energies lower than 31 GeV, FS is applied again; however due to the shrinking phase space the distributions in Fig.~\ref{fig:zfactors} converge to zero. For kaons the DDM cross section should decrease more rapidly at low energies due to strangeness threshold effects, but the impact of the very-low-energy kaon yields on the atmospheric fluxes is negligible.

The second model assumption is isospin symmetry (see, \eg{}, Ref.~\cite{Gaisser:2016cr}), which is required to relate particle yields between proton and neutron projectiles, and between $\pi^-$ and $\pi^+$ projectiles, respectively. No significant deviation is known for forward longitudinal spectra at relevant energies, except those related to different feed-down corrections and the definition of {\it stable} particles. \revise{The $K^0_\text{L,S}$ yields are calculated from the isospin relation $(K^+ + K^-)/2$, which is a valid approximation for the carbon target} \cite{Abgrall:2015hmv} but not necessarily for $pp$ interactions \cite{Anticic:2010yg}. In the present version of the \ddm{}, isospin symmetry has not been applied to antibaryon distributions since antibaryon interactions negligibly contribute to inclusive flux calculations, contrary to air showers. 

A third assumption is made for the production (inelastic) interaction cross sections, which are taken from \dpmjet{-III 19.1}. These have been recently updated using LHC measurements \cite{Fedynitch:2015kcn} and compared to proton-carbon measurements in Ref.~\cite{Bhatt:2020dxk}. The impact of the interaction lengths on inclusive fluxes is small with respect to the errors of the cross-section measurements. The differences between using carbon and air targets were studied using different event generators and found to be negligible ($<1\%$). Additional, minor simplifications originate from \mceq{} as cascade code, such as the superposition of primary projectile nuclei.

\subsection{Impact of individual channels}

\revise{As one may expect from the model comparisons in 
\figs~\ref{fig:pC_data_mesons} and \ref{fig:zfactors}, the modified $\pi^+$ and $K^-$ yields have the largest impact on inclusive flux calculations, since these} \dpmjet{} and \eposlhc{} both consistently underestimate or overestimate yields. Figure \ref{fig:impact_channels} helps to assess the differences of the new model with respect to these previous calculations based on event generators. The $\pi^+$ yields have the largest extrapolation uncertainty and (as discussed later) dominate the uncertainty estimation. At low energies baryons have an impact on muon fluxes since these can effectively change the average production depth, which is relevant for unstable particles.

The substantial change in low-energy muons due to the pion and proton yields is reflected in the sub-GeV--GeV neutrino fluxes (upper left panel of Fig.~\ref{fig:impact_channels}). The impact on the low-energy $\nu_\text{e}/\bar{\nu}_\text{e}$ ratio is compensated once all channels are simultaneously active. Except for the $\pi^+$ yields, the descriptions of fluxes by the event generators are satisfactory. For the models \sibyll{-2.3d} and \qgsjet{-II-04} (not shown in Fig.~\ref{fig:impact_channels}) the differences are slightly larger, which is mainly related to $\pi^+$ and $K^-$. Compared to \eposlhc{}, the \ddm{} produces less baryons, and for mesons larger differences are observed for $\pi^-$ and $K^+$.

\begin{figure*}
  \centering
  \includegraphics[width=0.92\textwidth]{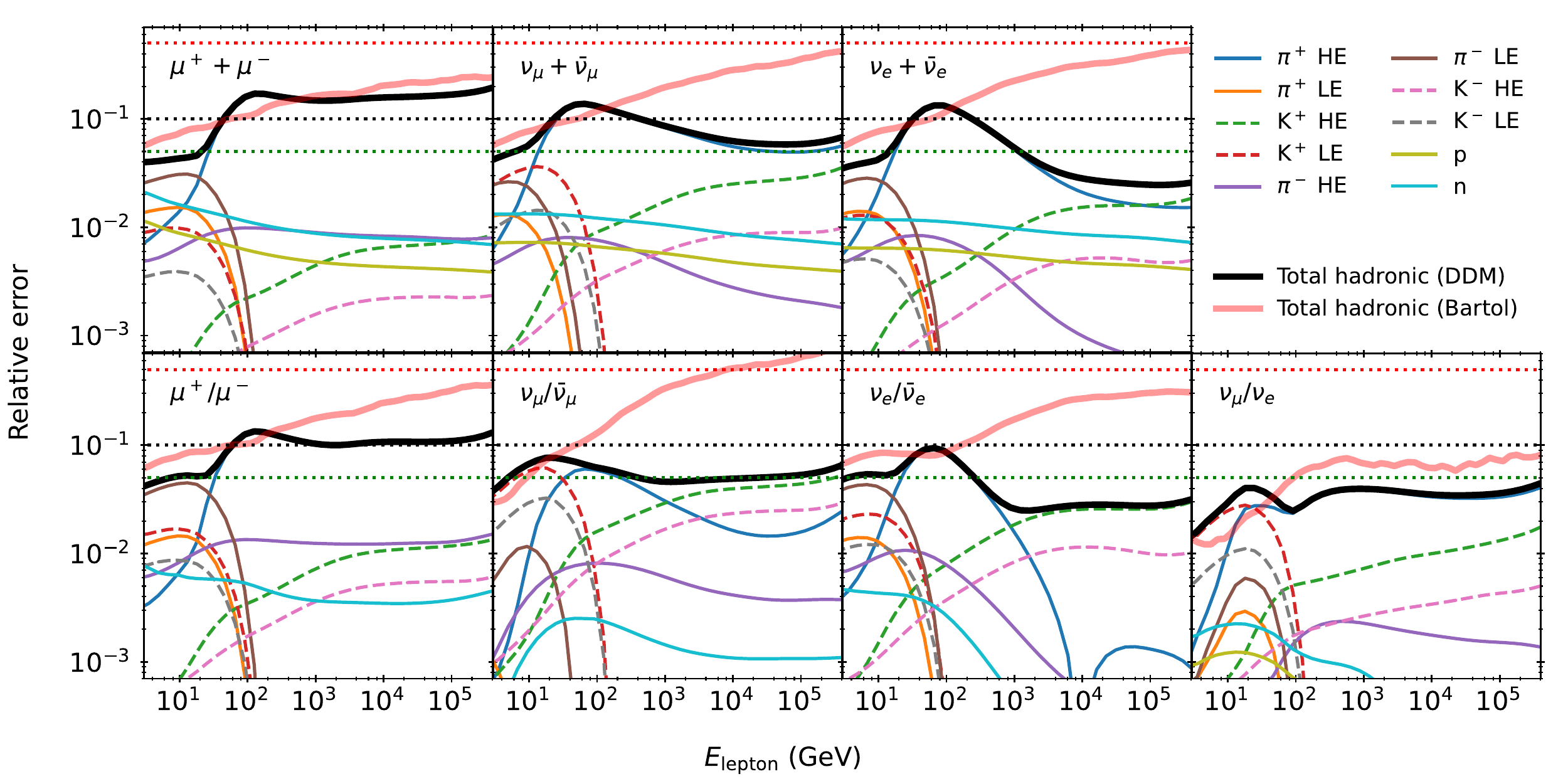}
  \caption{Estimated hadronic uncertainties on the conventional fluxes from \ddm{} compared to that from the BES. The detailed break-down into individual components of BES is shown in \fig~\ref{fig:bartol_uncertainties} within Appendix \ref{app:bartol_scheme}. The green, black and red dotted lines represent 5\%, 10\%, and 50\% uncertainty to guide the eye. The HE (high-energy) and LE (low-energy) labels correspond to uncertainties originating from the 158 GeV and the 31 GeV data, respectively. The cosmic ray flux uncertainty is not included.}
  \label{fig:uncertainties_ddm_vs_Barr}
\end{figure*}
\section{Uncertainties of inclusive fluxes}

At atmospheric lepton energies above $\sim$\,50 GeV, the dominant uncertainty is clearly the $\pi^+$ production measurement at 158 GeV (thin blue curve in \fig~\ref{fig:uncertainties_ddm_vs_Barr}). The apparent bump in the total uncertainty (thick black curve) is related to the threshold of the high-energy pion data set in \ddm{} and the gradual transition to the kaon-decay-dominated energy range (see, \eg{}, Ref.~\cite{Fedynitch:2018cbl}). The \textit{Bartol error scheme}\footnote{This scheme is sometimes called \textit{Barr parameters}. The implementation in \mceq{} is described in Appendix \ref{app:bartol_scheme}} (BES) \cite{Barr:2006it} produces larger uncertainties at high energies, since it assumes a 40\% uncertainty for $K^\pm$ production. An additional energy-dependent extrapolation uncertainty in the BES generates the steady rise of the light-red bands. Since we used NA49 proton-proton data at 158 GeV to model charged kaons, the hadronic uncertainties from the \ddm{} are much smaller despite the additional errors from the model-dependent extrapolation. \revise{At neutrino energies $\gtrsim1$\, TeV, the leading 25\%-uncertainty in our scheme would stem from the cosmic-ray fluxes \cite{Fedynitch_ICRC2017}. Above 100 TeV the uncertainties are dominated by the contribution of the poorly know forward charm yields}; see \eg{}, Ref.~\cite{Benzke:2017yjn}.

Below 10 GeV, the \ddm{} reduces the uncertainties compared to the Bartol scheme due to the phase-space coverage of the more recent NA61 measurement at $p_\text{Beam}=31$ GeV. The uncertainties for low-energy inclusive muons are dominated by the proton and neutron yields. To understand this effect one may consider that a higher elasticity in baryon interactions results in higher-energy secondary baryons that can produce more secondaries further downstream of the cascade. If muons are produced closer to the ground, fewer decay in flight. For this reason, the impact of baryons is weaker for neutrinos. For particle ratios the impact from baryons cancels out as expected.

The uncertainty for $\nu_\mu/\bar{\nu}_\mu$ at energies relevant for atmospheric neutrino oscillations is dominated by the uncertainty of the 31 GeV kaon data. Since it contributes very little to muon observables, obtaining better constraints on $\nu_\mu/\bar{\nu}_\mu$ through calibration with muon spectrometer measurements such as in Ref.~\cite{Honda:2019ymh} or Ref.~\cite{Yanez:2019bnw} is not feasible. Therefore, a further reduction of the hadronic uncertainty below 5\% requires a higher-statistics fixed-target measurement. For $\nu_\text{e}/\bar{\nu}_\text{e}$, the prospects for muon calibration are better since pion uncertainties can be constrained by muon flux and charge ratio measurements.

\section{Fluxes and charge ratios from the \ddm{}}

\subsection{Muon flux and charge ratio}
\label{sec:muon_discussion}
\begin{figure*}
  \centering
  \includegraphics[width=.49\textwidth]{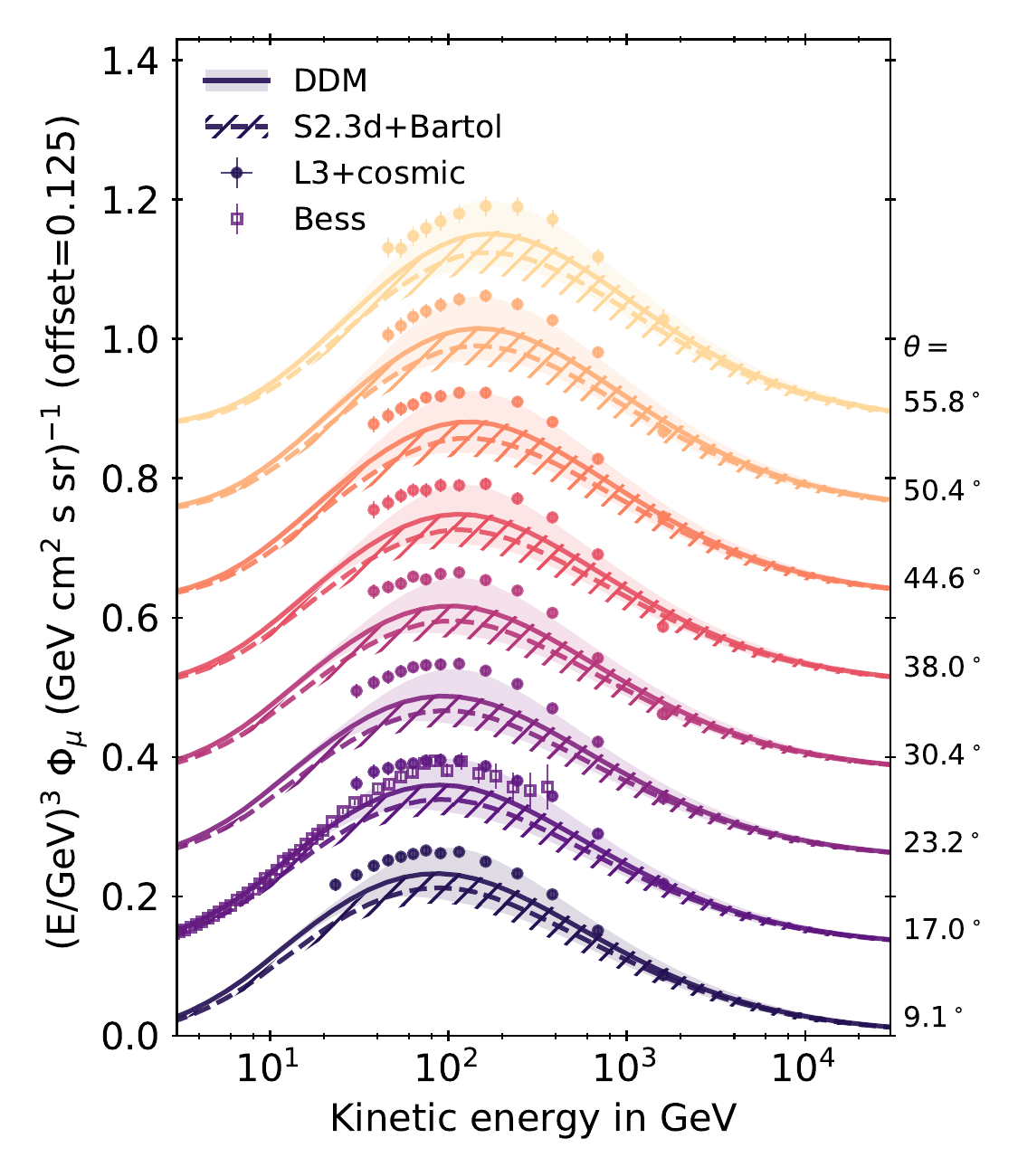}
  \includegraphics[width=.49\textwidth]{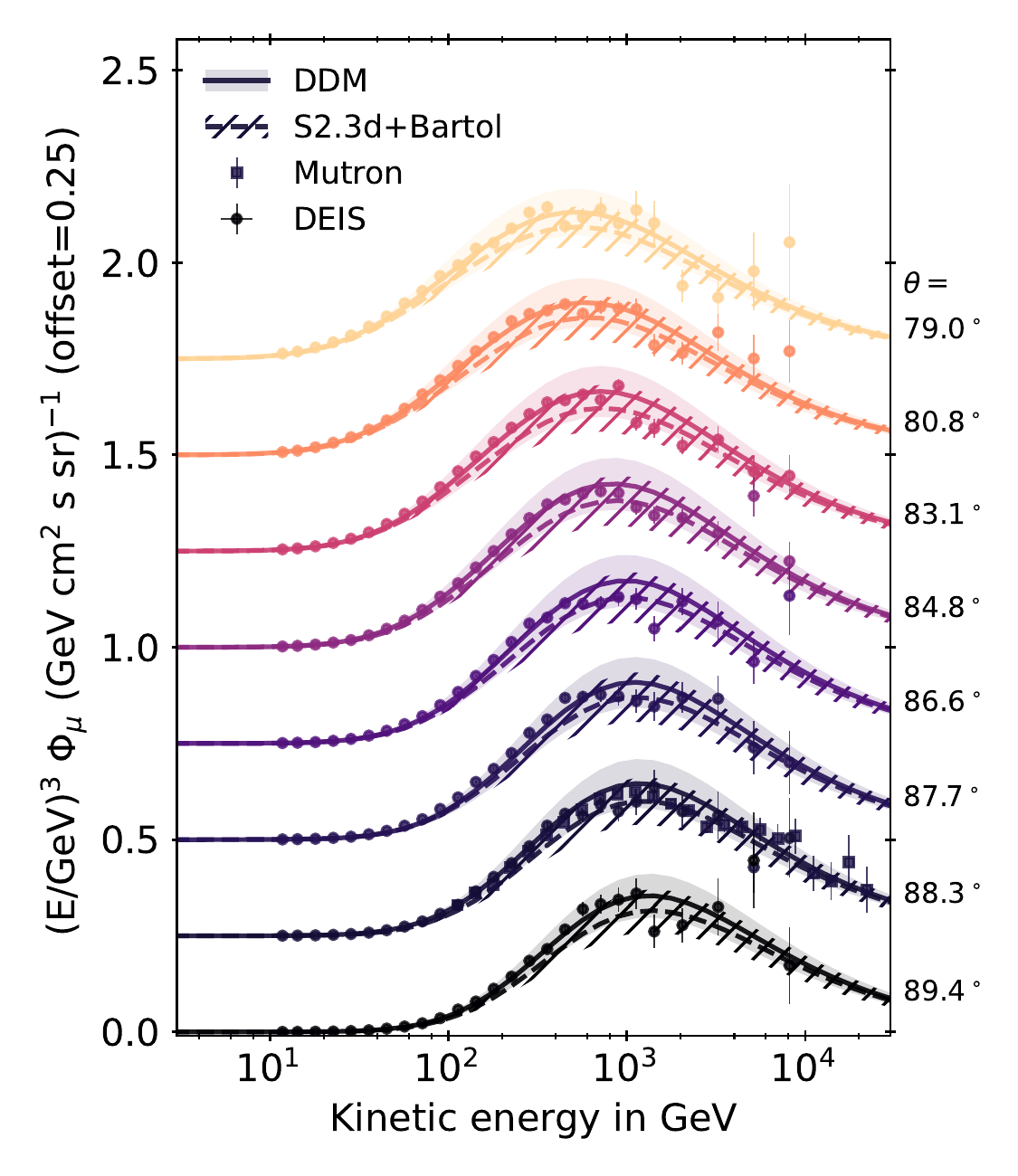}
  \includegraphics[width=.49\textwidth]{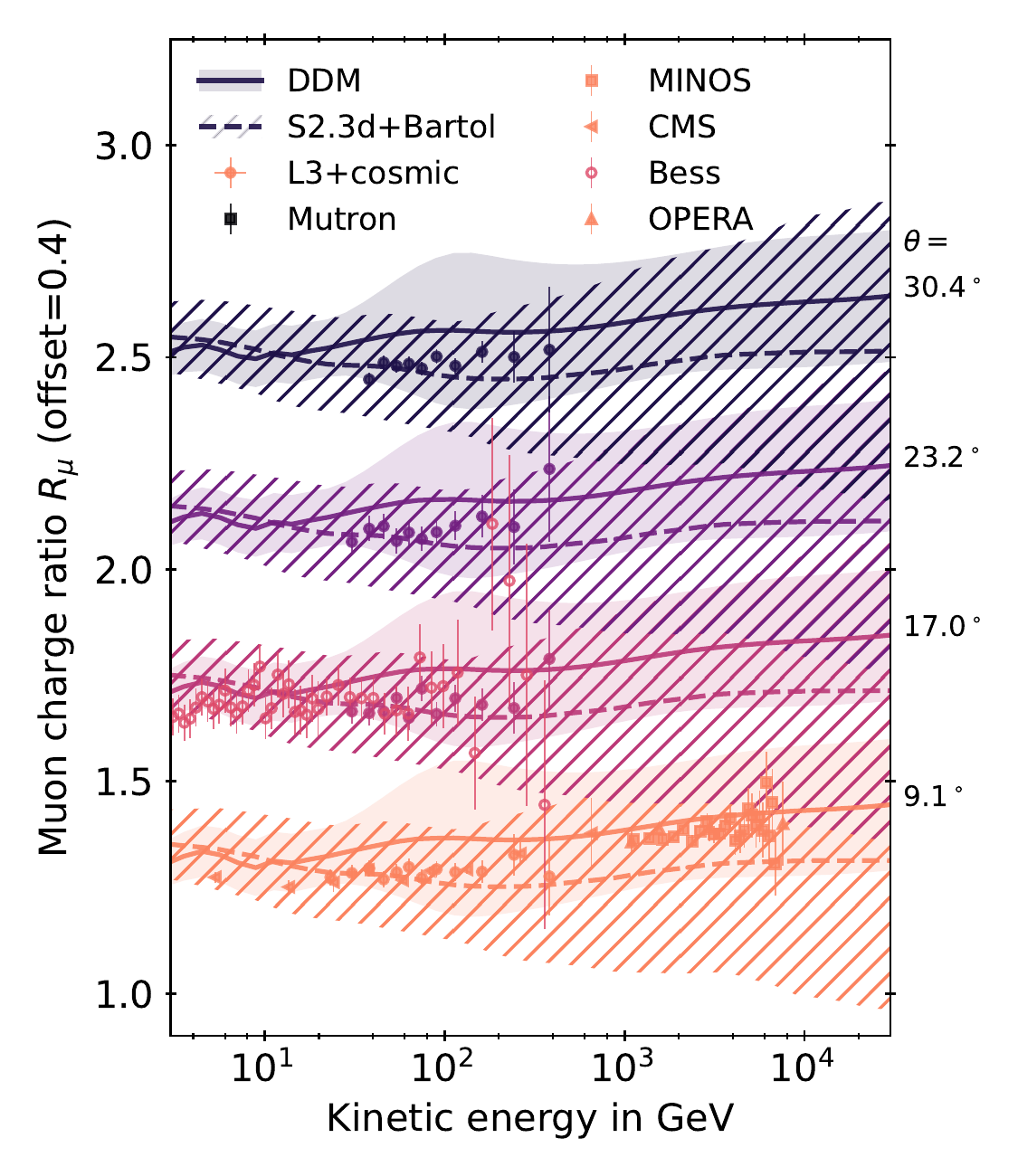}
  \includegraphics[width=.49\textwidth]{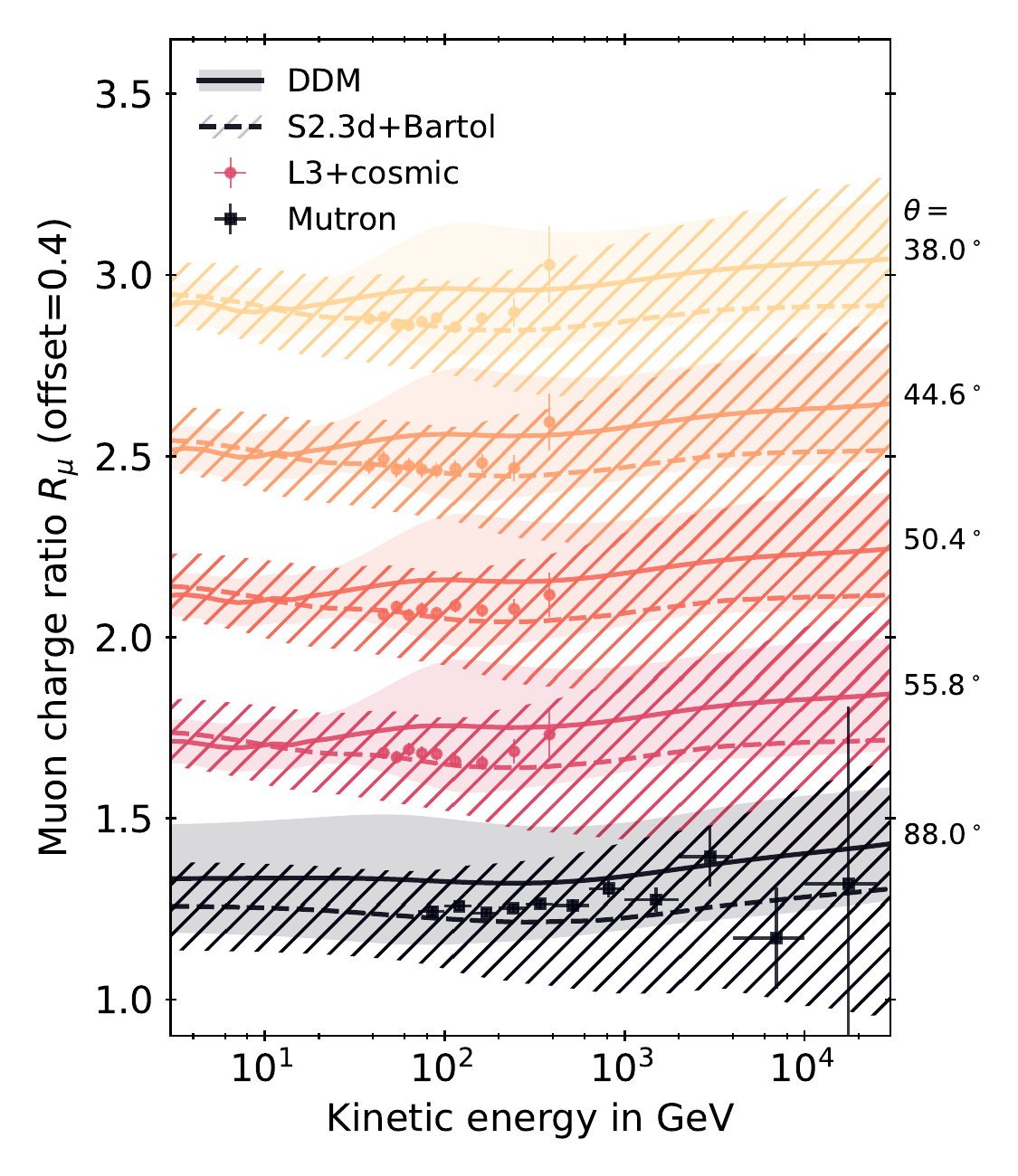}
  \caption{Inclusive muon fluxes (top panels) and charge ratios (bottom panels) at various zenith angles. As indicated in the $y$-axis labels, offsets have been applied for visual separation of curves above the lowest ones. The cosmic-ray flux model in both cases is GSF19 \cite{Schroder:2019agg}. Systematic uncertainties of the data \cite{Achard:2004ws,Haino:2004nq,Allkofer:1985ey,Matsuno:1984kq,Adamson:2007ww,Khachatryan:2010mw,Agafonova:2014mzx} are not shown.}
  \label{fig:atm_mu}
\end{figure*}

In combination with \mceq{}, the \ddm{} can be directly applied to calculations of atmospheric fluxes and uncertainties using data-driven models for hadronic interactions and the cosmic-ray spectrum. Here, we use a more recent version of the Global Spline Fit (GSF19), which slightly changes the \sibyll{-2.3d} prediction compared to the results presented in Ref.~\cite{Fedynitch:2018cbl}.

Inclusive muons and the muon charge ratio are shown in \fig~\ref{fig:atm_mu}. \revise{The left panels show near-vertical and the right panels near-horizontal zenith angles, respectively.} For the near-vertical directions, the central flux predictions (solid curves) match the data up to 100 GeV without applying corrections for experimental systematics. Above 100 GeV, the data is within the $1\sigma$ uncertainty band but the center prediction is a few percent below the Bess data \cite{Haino:2004nq}. For L3+c \cite{Achard:2004ws}, the systematic uncertainties on the energy scale are not shown but can be expected to have a sufficient impact on the normalization to bring the data in line with the calculation. The near-horizontal TeV muons shown in the top right panel are well described by the \ddm{}. The remaining differences between the DDM prediction and the data will be addressed in more detail in an upcoming work \cite{Yanez:2019bnw}.

For the muon charge ratio, shown in the bottom panels of \fig~\ref{fig:atm_mu}, the central value of the \ddm{} + GSF combination is somewhat higher than the data but consistent with it within uncertainties. A small reduction of the $\pi^+$ yield within the range of the \ddm{} 1$\sigma$ uncertainty could improve the agreement of the central value at the cost of slightly more tension in the vertical fluxes. The most vertical zenith bin (orange, bottom left panel) includes data at TeV energies from MINOS and OPERA that more sensibly probe the kaon charge ratio. One of the problems in the interpretation of the data is that it has been unfolded to equivalent surface energies under a simplified assumption for the angular scaling $\Phi_\mu(\theta) = \Phi_\mu(0^\circ)/\cos{\theta^*}$, which is only approximately valid at small angles $<30^\circ$ (cf.~Fig.~7 in Ref.~\cite{Fedynitch:2021ima}). Flux  ``calibration'' applications (such as those in Refs.~\cite{Yanez:2019bnw,Sanuki:2006yd}) would profit from underground muon rates measured as a function of the zenith angle and the slant depth in kilometer water equivalent (km.w.e.), even if only a few bins are populated.

Without applying larger systematic shifts to the muon data, the general conclusion is that the calculated flux needs to be less than $10\%$ higher between $\SIrange[range-phrase=-, range-units=single]{100}{200}{GeV}$ to match Bess. This difference is absorbed by the (conservative) 1$\sigma$ bands of the \ddm{} model. The remaining main source of uncertainty are the cosmic ray proton and helium fluxes in the energy range between 100 GeV and a few tens of TeV. These should be well constrained by the space-borne detectors AMS-02 \cite{AMS:2015tnn}, CALET \cite{CALET:2019bmh}, and DAMPE \cite{DAMPE:2019gys}. There are, however, some existing systematic differences between these data \cite{Marrocchesi:2021kaf}, that may yield a few \% higher proton or helium fluxes above 200 GeV with some additional softening above $\sim 20$ TeV (to keep higher-energy fluxes at about the same value). An inconsistency between calculations and muon flux measurements has been previously discussed in the literature \cite{Lagutin:2004ka}. Our current result indicates a similar trend quantitatively by using data-driven models. However, due to the complex nature of the cosmic ray measurement systematics and the hadronic model uncertainties, a true disagreement may not exist.

\subsection{Muon and electron neutrino fluxes}
\begin{figure}
  \centering
  \includegraphics[width=.85\columnwidth]{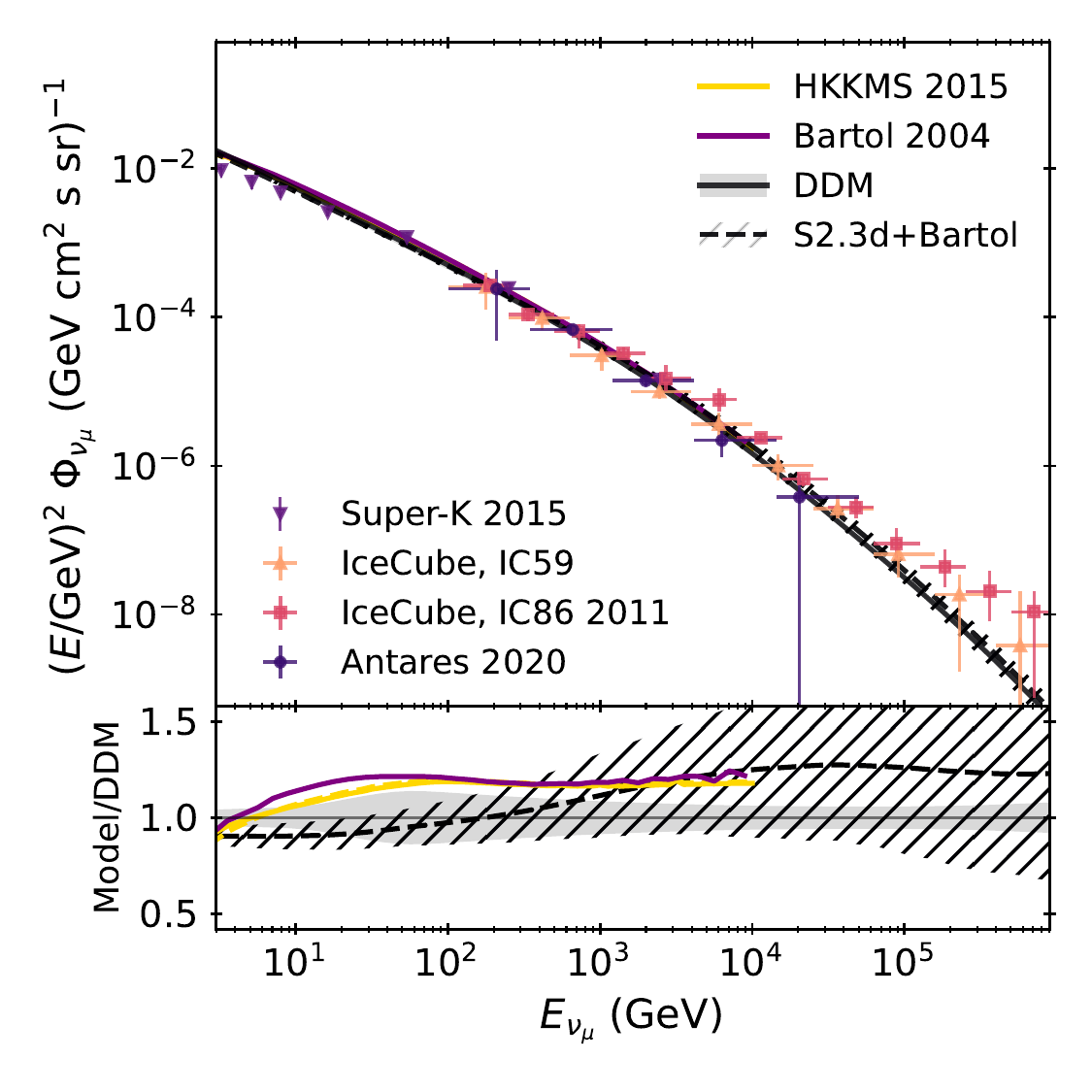}
  \includegraphics[width=.85\columnwidth]{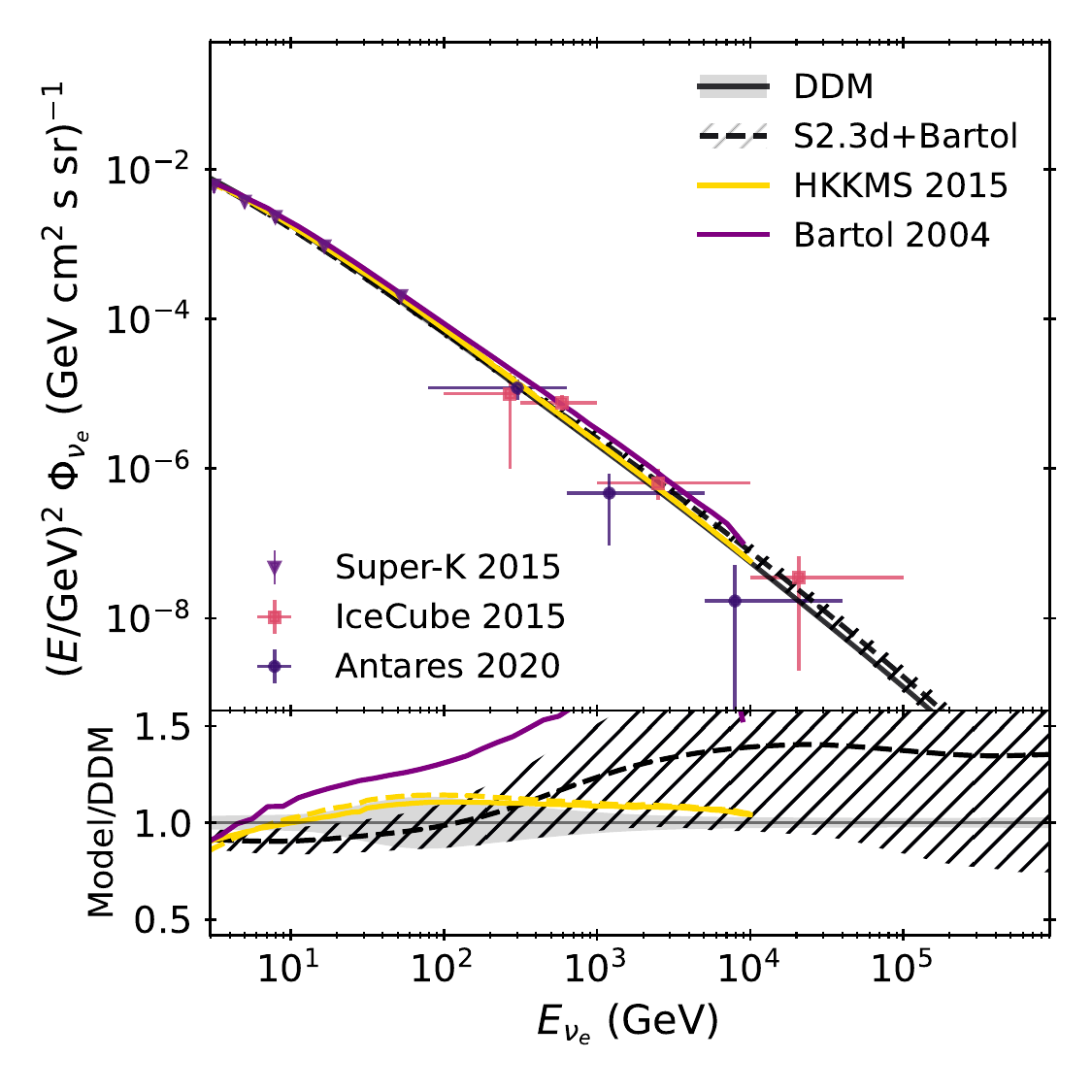}
  \caption{Conventional atmospheric neutrino fluxes averaged over zenith angles $\theta$. The prompt component is not shown. The data \cite{ANTARES:2021cwc,Richard:2015aua,Aartsen:2014qna,Aartsen:2015xup} covers a wide energy range from MeV to almost PeV energies. The $\nu_\mu$ flux data from Super-K below 10 GeV is affected by $\nu_\mu$ disappearance, which is not included in the calculations. The \ddm{} result is compatible with \sibyll{-2.3c} from a previous work \cite{Fedynitch:2018cbl,Fedynitch_ICRC2017} within errors above 1 GeV. However, the HKKMS and the Bartol calculations are outside of the \ddm{} error band for $\nu_\mu$. For HKKMS the dashed line is the average flux at Kamioka whereas the solid line is computed for the South Pole, giving an estimate of the energy range impacted by geomagnetic and 3D effects.}
  \label{fig:atm_nu_fluxes}
\end{figure}

The muon and electron neutrino fluxes are compared to some reference calculations and data in \fig~\ref{fig:atm_nu_fluxes}. For muon neutrinos (upper panels), the new model is compatible with the data within uncertainties. At the highest energies, two bins of the IceCube IC86 measurement are higher than the calculation at less than $2\sigma$, indicating either a contamination of atmospheric fluxes by astrophysical or prompt neutrinos, or that the uncertainties of the data could be underestimated. The ANTARES \cite{ANTARES:2021cwc} and IC59 \cite{Aartsen:2014qna} measurements are well described. The Super-K measurement below 30 GeV is not corrected for $\nu_\mu$ disappearance and cannot be directly compared to the models. Compared to the HKKMS (and Bartol) calculations some disagreement is expected since HKKMS has been tuned to fit muon measurements. A deficit of several \% has been found (see \Sec~\ref{sec:muon_discussion}) for the muon fluxes, which is expected to translate directly into $\nu_\mu$ fluxes at $E_{\nu_\mu}<100$ GeV. \revise{Therefore, finding the HKKMS calculation to lie  $\sim20$\% above our prediction is larger than expected.} The description of fluxes in the TeV range by \mceq{} and the \ddm{} has been recently studied for underground muon intensities \cite{Fedynitch:2021ima}, and found to be in good agreement with vertical intensity data, and the error estimation of muon fluxes in the \ddm{} has been demonstrated to be realistic.

Below a few \revise{tens} GeV, the \mceq{} calculations with the \ddm{} agree better with Bartol and HKKMS fluxes compared to previous estimates that use \dpmjet{} as the low-energy interaction model (the \sibyll{-2.3d}+Bartol curves are calculated using \dpmjet{} below 80 GeV projectile energy). The two factors equally contributing to this result are the \ddm{} and the update from the original GSF to the newer GSF19 fit. This energy range is not the main focus of the present \ddm{} model and a more complete result will be obtained by including 3D and geomagnetic effects. It would also be important to use the HARP data at the lowest energies since the assumption of scaling of hadronic yields below 31 GeV in the \ddm{} is \revise{invalid for neutrino fluxes below $\sim3$ GeV}. Investigating these aspects is beyond the scope of this work and requires a dedicated low-energy calculation.

The comparison between the calculations for electron neutrinos (lower panel of \fig~\ref{fig:atm_nu_fluxes}) shows a similar result. The Super-K data is now well described by the \ddm{} + GSF19 calculation. The high-energy data from ANTARES is compatible with our result; however, it is also notably lower than the IceCube result and all of the calculations at its asymmetric bin centers (in particular, when scaled with $E^3$). From the discussion of the kaon $Z$ factors alone (\fig~\ref{fig:zfactors}), the \ddm{} should be expected to be significantly ($\sim40\%$) lower than the HKKMS calculation, but some of this difference is compensated by the cosmic-ray spectrum. \revise{The $\nu_e$ fluxes from the HKKMS model are 10--15\% higher than ours, and partis of the spectrum are consistent with our estimated error band,} in particular below a few tens of GeV, where $\nu_e$'s mostly originate from muon decays.

In the comparison between \sibyll{-2.3d} and \ddm{} within \mceq{}, differences occur at high energies, where the less abundant charged kaon component of the \ddm{} manifests as a shift in the neutrino spectral index. Within errors both models are compatible, but the \ddm{} calculation has significantly smaller errors.

\subsection{Neutrino ratios}
\begin{figure}
  \centering
  \includegraphics[width=.85\columnwidth]{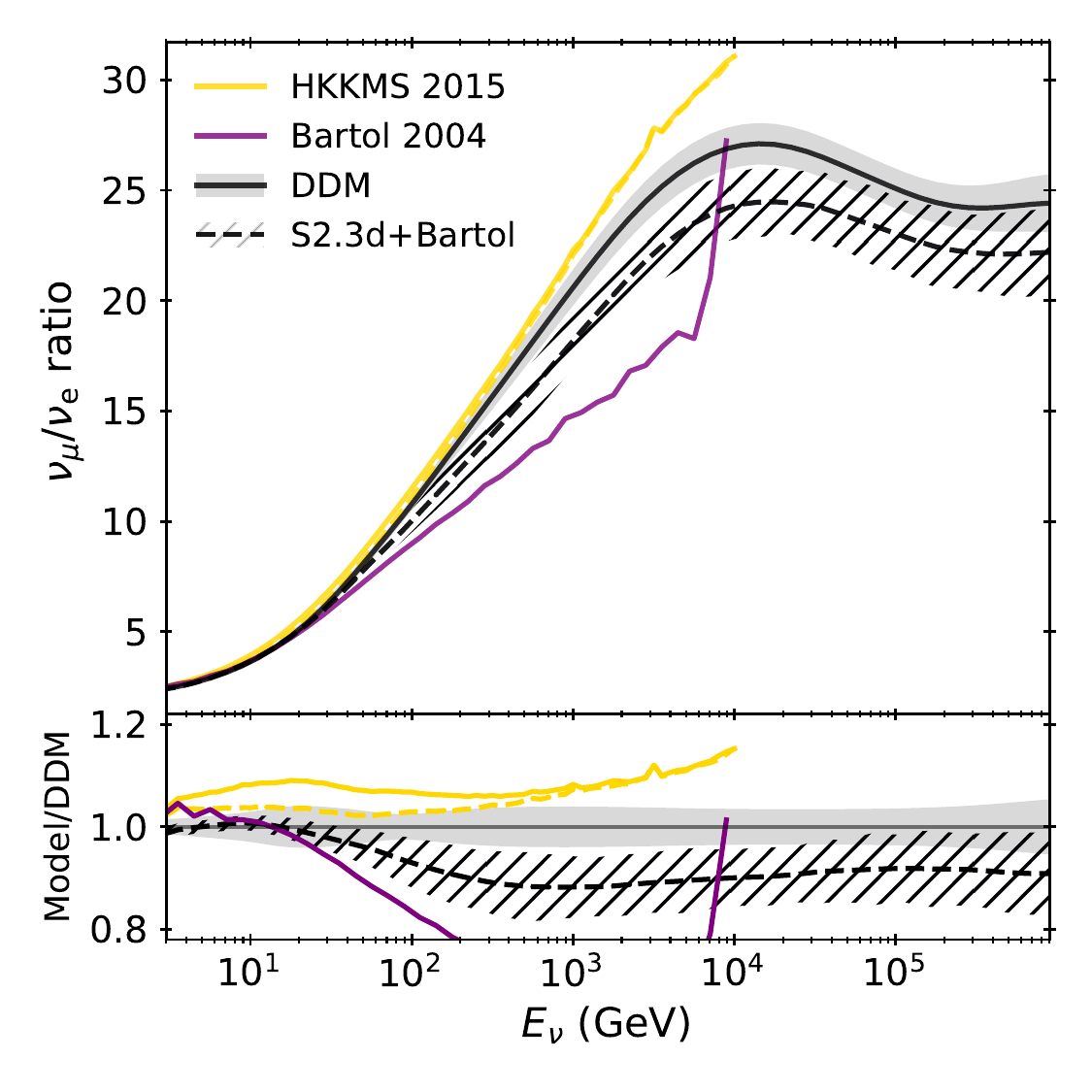}
  \caption{Conventional, zenith-averaged flavor ratio $(\nu_\mu+\bar{\nu}_\mu)/(\nu_\text{e}+\bar{\nu}_\text{e})$. The deviation above 10 GeV between the \mceq{} and 3D calculations is related to the kaon content predicted by the models (compare with \figu{zfactors}). More forward neutral and charged kaons yield more electron neutrinos at intermediate energies. At very high energies, muon and electron neutrinos \revise{both scale proportionally to the charged kaon content leading to the flattening of their ratio.}}
  \label{fig:atm_flavor_ratios}
\end{figure}

\begin{figure}
  \centering
  \includegraphics[width=.85\columnwidth]{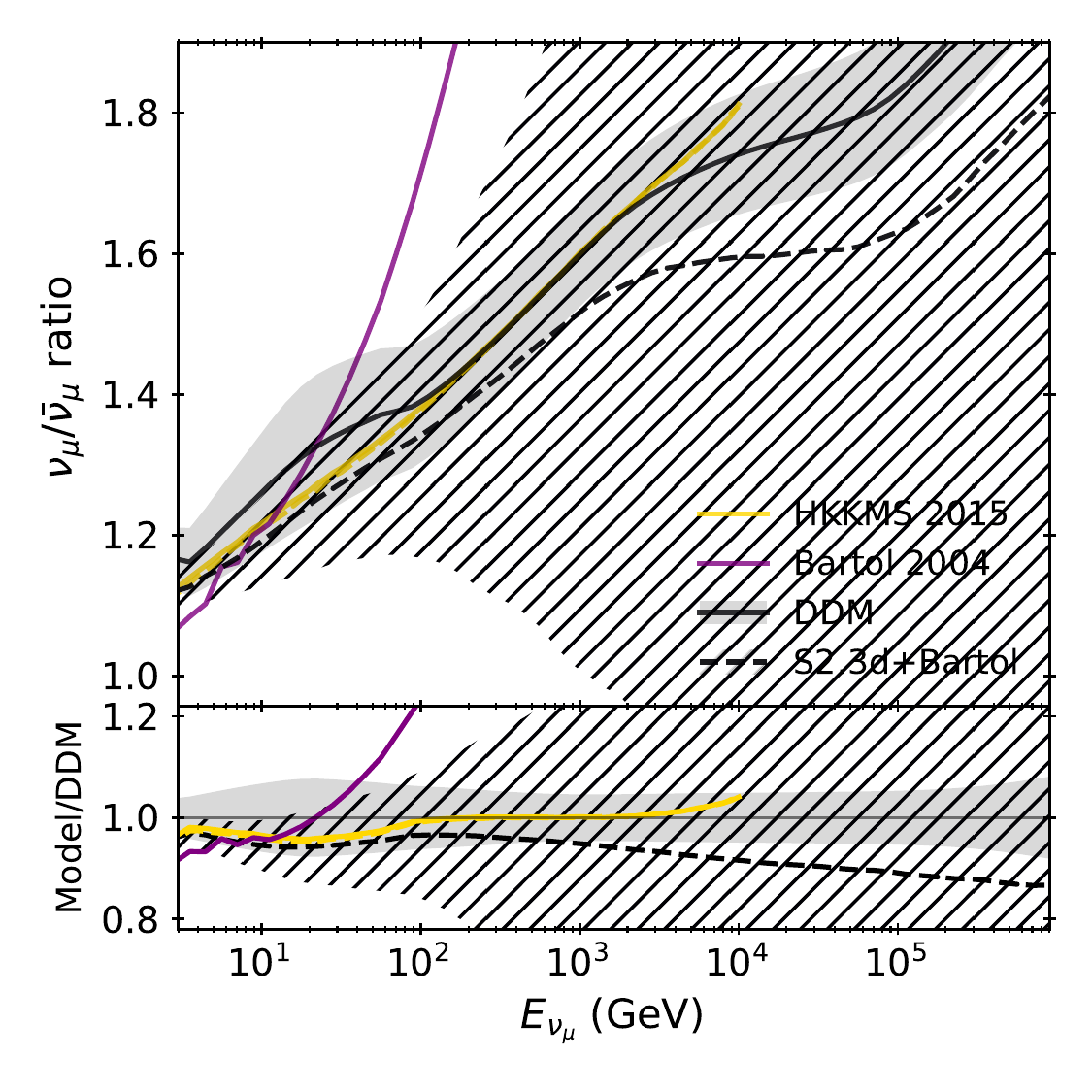}
  \includegraphics[width=.85\columnwidth]{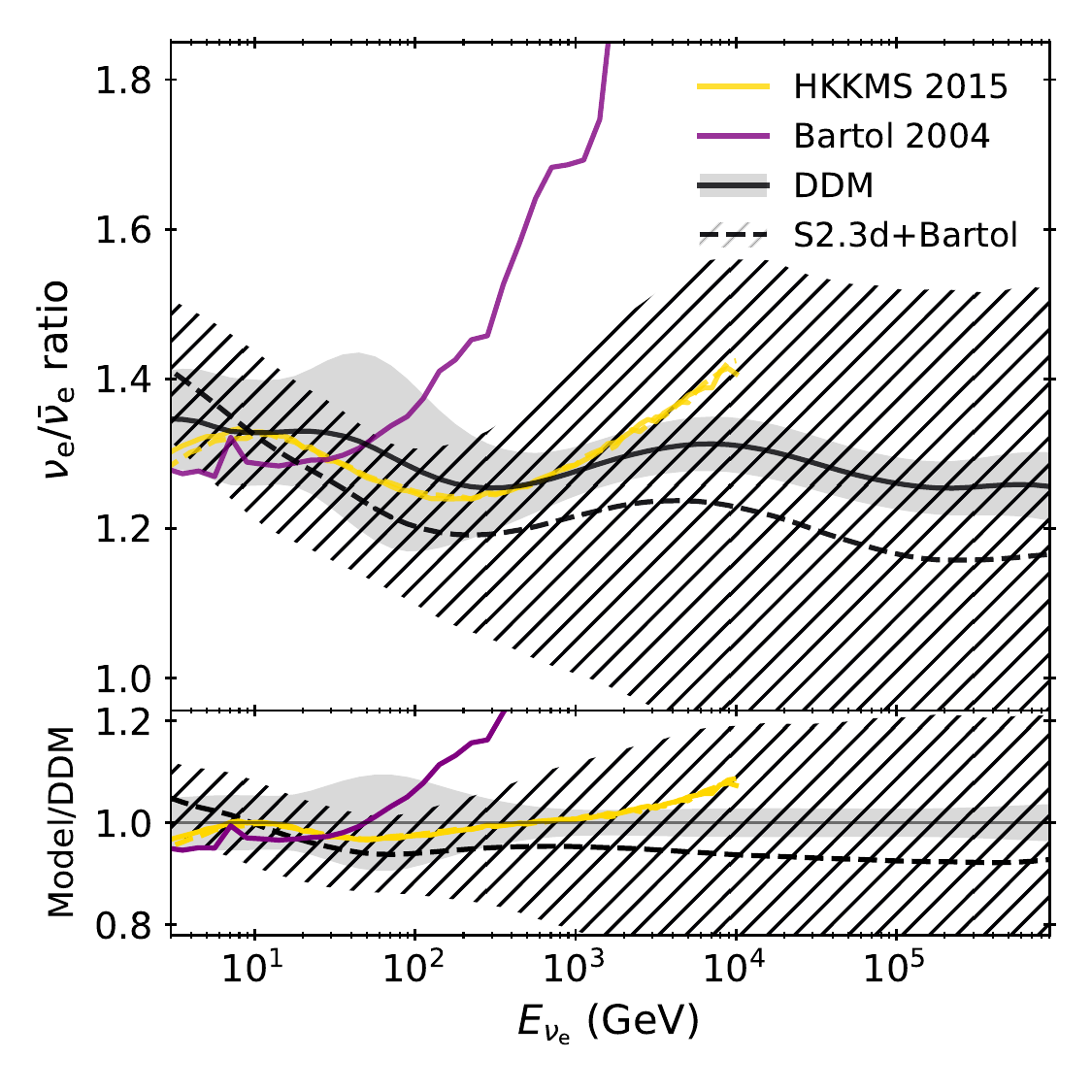}
  \caption{Conventional, zenith-averaged neutrino-antineutrino ratios. The \ddm{} uncertainties are significantly smaller than those estimated with the BES (hatched). The high-energy error in the BES is mainly affected by kaon uncertainties of $\sim$\,30\% and an additive {\it ad hoc} extrapolation uncertainty. The recent calculations agree within uncertainties over a large energy range.}
  \label{fig:atm_nu_ratios}
\end{figure}

Larger differences that can be experimentally relevant are observed in the flavor ratios, as shown in \figu{atm_flavor_ratios}. At energies below 5 GeV the flavor ratio can be affected by geometrical limitations of the one-dimensional approach in \mceq{}, whereas at high energies it is affected by the energy dependence of the $K$/$\pi$ ratio. Compared to the HKKMS model, \mceq{} shows an almost constant offset of $\sim10$\% due to lower muon neutrino fluxes, visible in the lower panel of \figu{atm_flavor_ratios}.

The neutrino-antineutrino ratio calculations in \figu{atm_nu_ratios} are compatible within uncertainties with the exception of the Bartol flux which suffers from large $K^+$ production. The most striking difference is the reduction of the hadronic uncertainty in the \ddm{} with respect to the BES, which dramatically shrinks above TeV energies. This change is mainly driven by smaller uncertainties on charged kaons, and by the absence of an \textit{ad hoc} extrapolation uncertainty in the \ddm{}. The \ddm{} prediction may be perceived to be optimistic, but the comparisons with the highest-energy muon fluxes and charge ratios in \figu{atm_mu}, as well as with the underground intensities in Ref.~\cite{Fedynitch:2021ima}, show that the uncertainty bands are not too narrow.

\section{Conclusion}
\label{sec:discussion}

The \ddm{} is a basic and relatively simple model of inclusive hadron production yields for interactions of protons or pions with light nuclei. It integrates the double-differential data in $(p, \theta)$  or $(x_\text{F}, p_\perp)$ taken at fixed-target experiments, \revise{propagating the uncertainties} to a single-differential cross section in $\xl$, which is \revise{an adequate choice} for one-dimensional cascade calculations with \mceq{}. The \ddm{} is cross-checked against atmospheric muon data and other calculations, and showed results are similar to or better than calculations based on traditional hadronic interaction models. The \ddm{} simplifies the assessment of systematic or theoretical errors on atmospheric fluxes since variations to the yields of hadrons are constrained within regions allowed by the data from accelerators. Due to these physical ``priors'', the \ddm{} is an optimal choice as a baseline flux model in neutrino-telescopes data analyses that struggle with quantifying the flux uncertainties. \revise{Percent-level-precision atmospheric lepton fluxes could be achieved using tighter, data-driven} constraints from a calibration with inclusive atmospheric surface or deep-underground muons \revise{with} the \ddm{} as the baseline hadronic flux model.

\acknowledgments

We would like to thank Juan Pablo Ya\~nez, Tetiana Kozynets, and Alfredo Ferrari for helpful comments. A.F.\ acknowledges the hospitality within the group of Hiroyuki Sagawa at the ICRR, where he completed parts of this work as a JSPS International Research Fellow (JSPS KAKENHI Grant Number 19F19750).

\bibliography{bibliography}

\onecolumngrid
\begin{appendix}

  \section{The \textit{Bartol error scheme} (BES) and error propagation in MCEq}
  \label{app:bartol_scheme}
  To propagate errors for one of the models involved in the calculations with \mceq{}, we use numerically computed gradients.
  \begin{figure}
    \centering
    \includegraphics[width=0.9\textwidth]{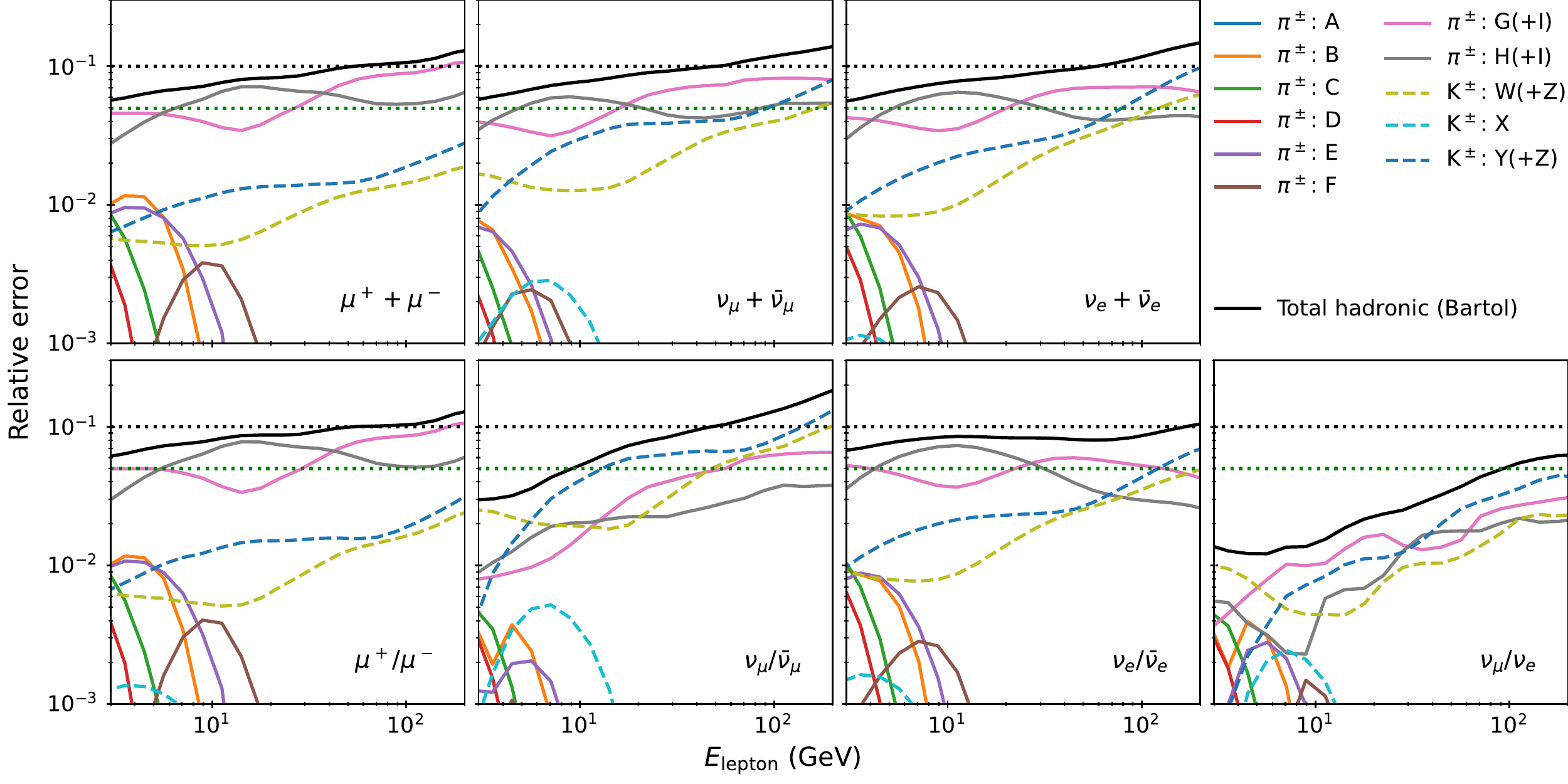}
    \includegraphics[width=0.9\textwidth]{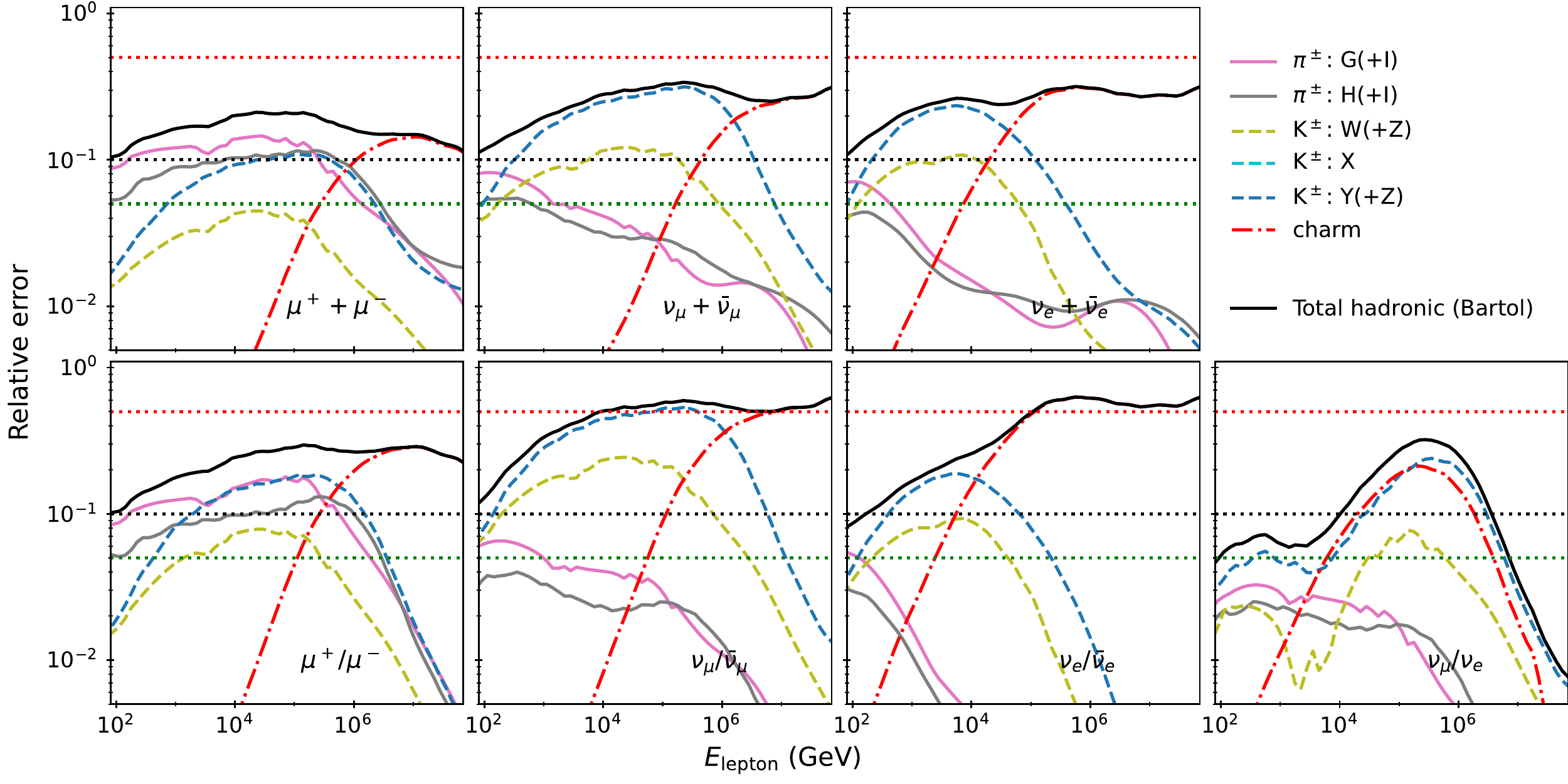}
    \caption{Uncertainties of atmospheric muon fluxes and ratios from the MCEq implementation of BES, at vertical zenith for muons and zenith-averages for neutrinos. The interaction model is \sibyll{-2.3d} and the primary model is GSF19. The letters A-Y correspond to phase space regions defined in Figs.~2 and 3 of \cite{Barr:2006it}. \revise{Equal uncertainties are assigned to particles of opposite charge and to the channels related by isospin symmetry}. The upper set of panels focuses on lower energies whereas the lower panels focus on high energies.}
    \label{fig:bartol_uncertainties}
  \end{figure}
  The BES \cite{Barr:2006it} is one of the more recent reference calculations for atmospheric neutrino flux uncertainties and \revise{lists relevant references in \Sec~II}. Errors are propagated using a one-dimensional Monte Carlo method starting from an assessment of accelerator data availability and precision. Similar to the approach taken in the present work, atmospheric muon data has not been used for the neutrino flux error estimation. To implement the BES scheme in MCEq, we create libraries of the neutrino flux gradients, computing them numerically via first order finite differences:
  \begin{equation}
    \frac{\partial \Phi_\ell(E_\ell, \theta, \mathcal{A}, \mathcal{B}, \dots)}{\partial \mathcal{A}} = \frac{
      \Phi_\ell(\dots, (1+\delta)\mathcal{A}, \dots) - \Phi_\ell(\dots, (1-\delta)\mathcal{A}, \dots)}{2\delta}.
  \end{equation}
  In the case of the BES, the calligraphic parameters modify the particle production cross sections from \equ{coupling_coeff} within $(E_\text{projectile}, E_\text{secondary}$)-ranges defined by each of the boxes shown in Figs.~2 and 3 of \cite{Barr:2006it}, whereas $\delta$ is some small number. These gradients are used to construct a Jacobian matrix and apply it in standard error propagation to project the uncertainties of each phase-space region on the lepton fluxes and ratios. The same technique can be applied to the propagation of uncertainties related to any of the models that take part in flux calculations such as the cosmic-ray nucleon flux or the atmospheric profile. In the case of the \ddm{}, gradients are computed with respect to the spline coefficients, which are obtained from the fit to the data (see \Sec~\ref{sec:uncertainty_parametrization}), and $\delta$ is their $1\sigma$ error. The error propagation is performed using the covariance matrix for the knots of each cross section fit from Figs.~\ref{fig:pC_data_mesons} and \ref{fig:pC_data_baryons}. In principle, the method can account for correlations between different particle species given sufficient data (e.g. a measurement of the $\pi^+/\pi^-$ ratio).

  The hadronic model uncertainties of conventional lepton fluxes, shown in Fig.~\ref{fig:bartol_uncertainties}, can be directly compared to the Figures 10 and 11 in Ref.~\cite{Barr:2006it}. Qualitatively, the schemes agree but there are some numerical differences probably related to significantly different kaon and pion yields between the interaction models. We verified that applying the BES to calculations made using other interaction models, such as \eposlhc{} or \qgsjet{}, yields very similar results. Also note, that this implementation of the BES is slightly different from what has been previously shown in Ref.~\cite{Fedynitch_ICRC2017} and \revise{to that} used by the IceCube Collaboration to estimate flux uncertainties. \revise{In these previous implementations, the extrapolation uncertainties (I and Z) were treated as independent parameters, which were quadratically summed with the other errors}. Instead, the extrapolation error is linearly summed with the error of the regions ($E_\text{projectile}>500$ GeV), and it also spans the entire phase space $\xl \geq 0$ in line with Ref.~\cite{Barr:2006it}, instead of $\xl \geq 0.1$ as in Ref.~\cite{Fedynitch_ICRC2017}.

  The BES is purely empirical construct and it is a solid attempt to conservatively parameterize the errors from the incomplete data coverage of the relevant particle production cross sections. However, one should not overlook the issues related to high energies, since at the time of construction the authors were focusing at Super-K energies rather than IceCube energies. The extrapolation errors (I and Z) have been assigned very conservatively, likely overestimating the true uncertainty of the $\nu/\bar{\nu}$ ratios. Since prompt fluxes have not been modeled, the BES can be used to extrapolate the uncertainty up to $\SIrange[range-phrase=-, range-units=single]{10}{100}{TeV}$, depending on the particle type, but not beyond that. One of the major issues is the weak connection of the hadronic interaction model used in the neutrino flux calculation that handles the interpolation between and extrapolation beyond the phase-space patches where data is available. A data-driven model, like \ddm{} or TARGET, can suffer from the inconsistency between partially overlapping data sets and thus end up outside of the quoted, purely experimental errors. At the same time, a physical or empirical model might lack sufficient parameters to describe all data within errors. A physical model, such as \dpmjet{}, has natural correlations between phase space patches that would lead to smaller errors, which are not handled by the BES. Finally, the subdivision into the different phase space patches is {\it ad hoc}. Therefore, one should regard the BES as a conservative estimate of the flux and ratio uncertainties.

  \section{Fits to NA61's pion-carbon data}
  \label{app:pi-c-fits}
  \begin{figure*}
    \centering
    \includegraphics[width=\textwidth]{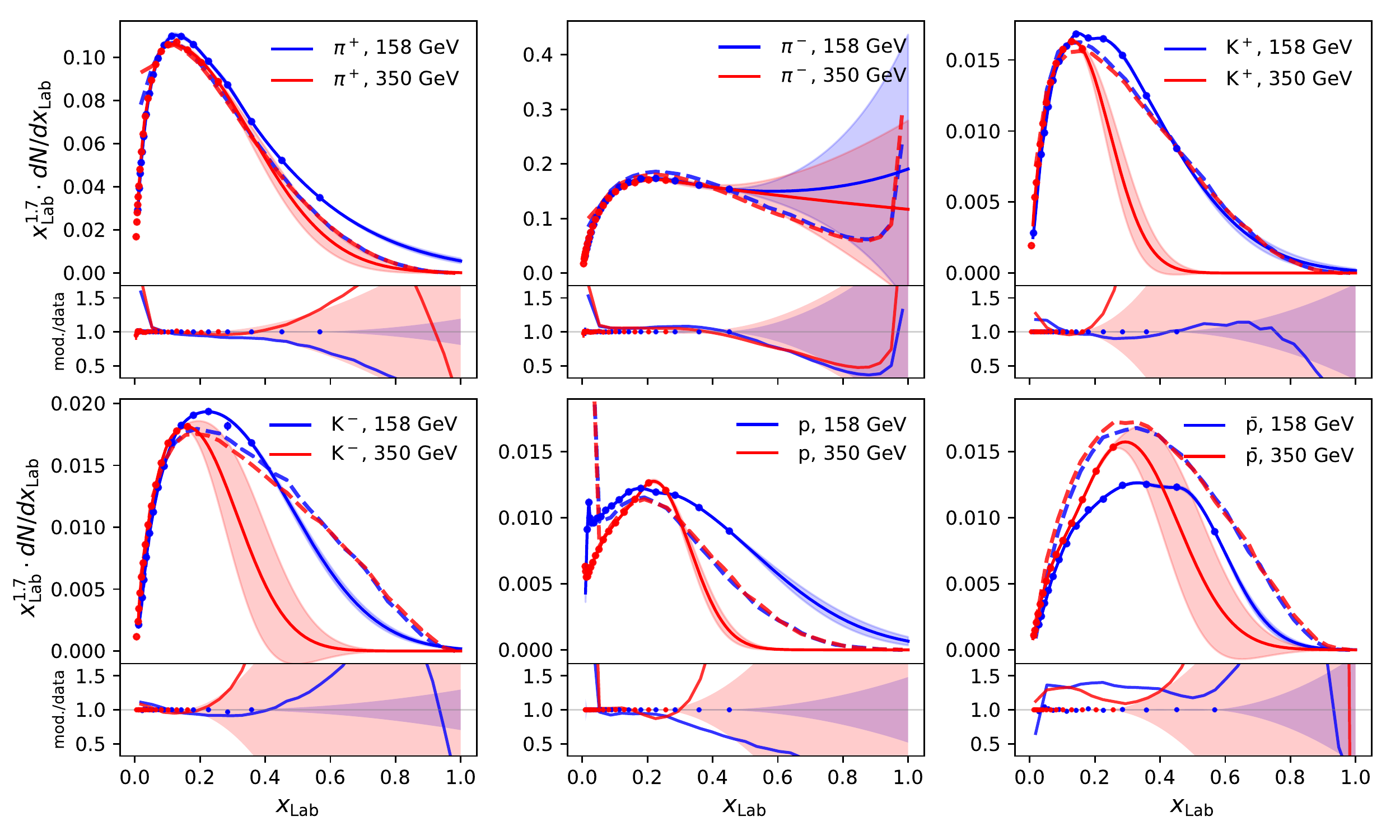}
    \caption{Inclusive production yields of
      different mesons in interactions of negative pions with a carbon target. The data are measured by
      the fixed target experiment NA61/SHINE (Table \ref{tab:na_data}). The error bars include
      systematic and statistical uncertainties. The solid curves show a spline fit to these data where the shaded band represents the uncertainty of the fit. For comparison the inclusive particle yields in pion-air collisions from \dpmjet{-III-19.1} are shown as dashed curves and have been obtained at the same projectile momentum.}
    \label{fig:piC_data}
  \end{figure*}

  \begin{figure*}
    \centering
    \includegraphics[width=.95\textwidth]{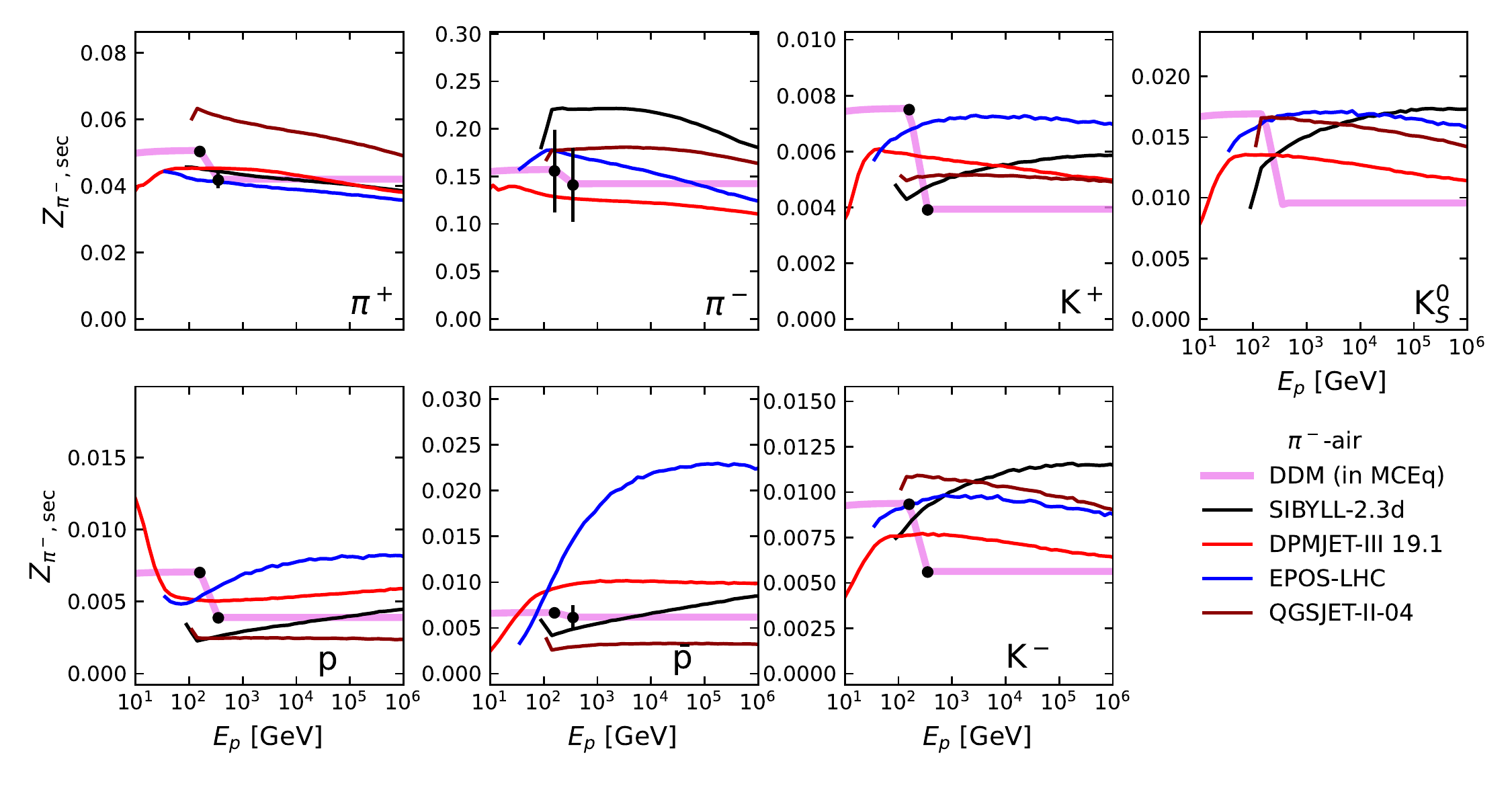}
    \caption{Energy-dependent spectrum-weighted moments ($Z$-factors) for $\pi^-$ air interactions and $\gamma = 2.7$ according to \equ{z-factor}.}
    \label{fig:zfactors-pions}
  \end{figure*}

  A model for the interactions of secondary pions (and kaons) with air is required to build a complete inclusive interaction model. For inclusive fluxes these interactions play a minor role but become important when modeling particle cascades initiated by individual cosmic rays (air showers). The large differences in the energy scaling between 158 GeV to 350 GeV in data are unexpected given the model predictions (other models predict similarly small differences between the two energies)\revise{, as shown in \fig~\ref{fig:piC_data}}. The difference in energy is only a factor two and particle multiplicities scale typically with $\sim\log{E_\text{projectile}}$. Within such a small interval and forward phase space, one expects almost perfect scaling of the order of what the dashed model curves show. Given the small quoted error of the NA61 data, it appears very challenging to explain the differences of many units of $\sigma$ from physics arguments. The small errors impact the fit quite significantly and lead to inconsistent results for the $Z$ factors in \fig~\ref{fig:zfactors-pions}, some of which lie many $\sigma$ apart. For future versions of the \ddm{}, one could combine the data into a single spectrum and inflate experimental errors until a single consistent fit emerges to quantify the true systematic uncertainty. Since the details of these secondary meson interactions are not important for inclusive flux calculations, this task will be left for a future work.

  \section{Extrapolation from proton to carbon target for charged kaons}
  \label{app:pp_pc_extrapolation}
  \begin{figure*}
    \centering
    \includegraphics[width=.45\textwidth]{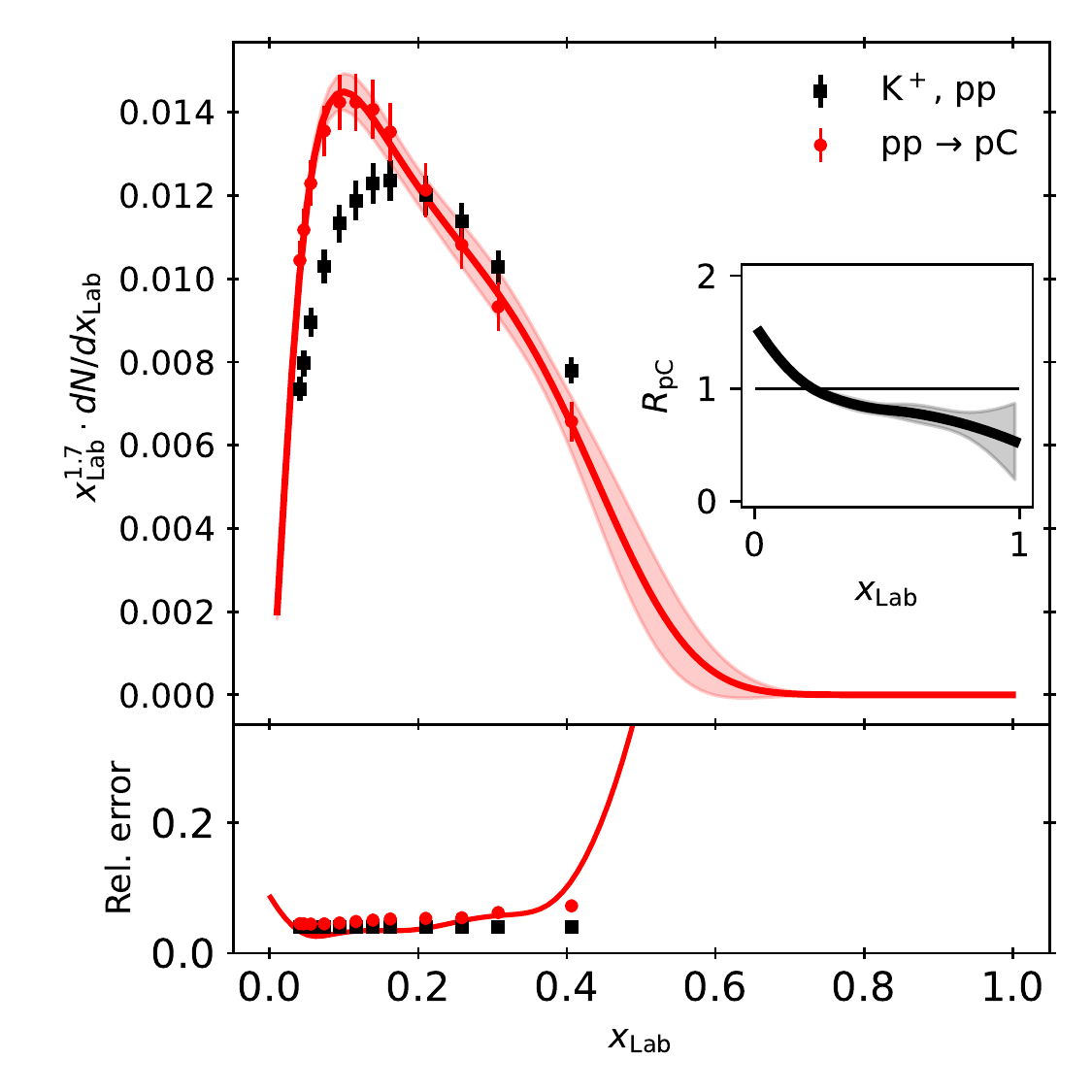}
    \includegraphics[width=.45\textwidth]{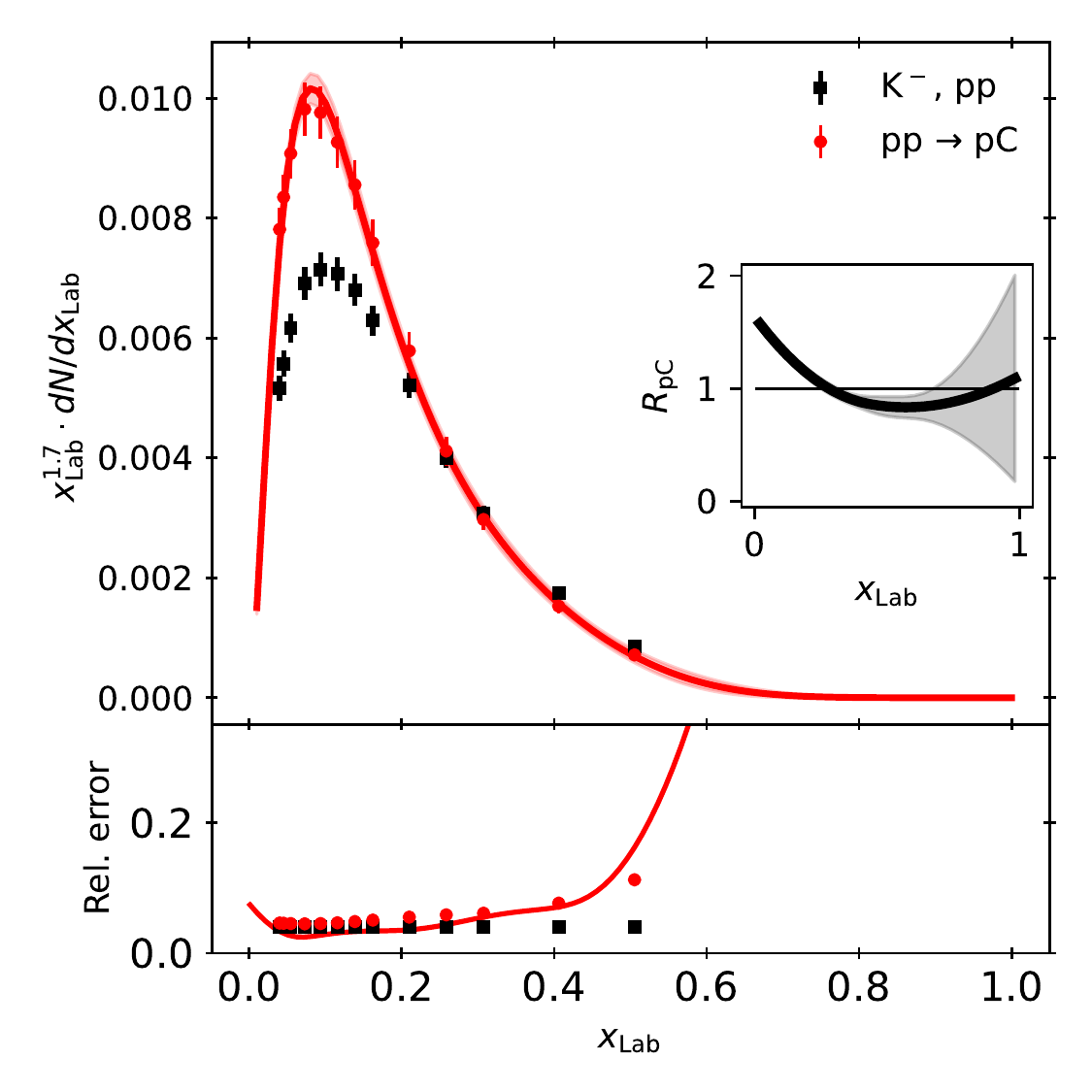}
    \caption{Extrapolation of differential kaon production cross section from NA49 proton-proton data at 158 GeV \cite{Anticic:2010yg} to proton-carbon. The data has been multiplied with the nuclear modification factor (shown as inset), which has been derived from an average of Monte Carlo simulations. The lower panel shows relative $1\sigma$ errors (error/data) and spline error divided by the best fit as a curve. 
    \label{fig:pp_pC_extrapolation}}
  \end{figure*}

  Neither NA49 nor NA61 has released any data for charged kaon production cross sections in proton-carbon collisions at 158 GeV. Since kaons play a major role for the production of neutrinos they cannot be ignored in the \ddm{}. The approach taken here is the extrapolation of kaon yields measured in proton-proton interactions to proton-carbon using a set of Monte Carlo interaction models. The mean and error of the nuclear modification factor 
  \begin{equation}
  	R_p\text{C}(\xl) = \frac{\rmd N_{p\text{C} \to K^+}}{\rmd \xl}/\frac{\rmd N_{pp \to K^+}}{\rmd \xl}
  \end{equation}
  is computed using an average of the predictions most recent versions of the \sibyll{}, \dpmjet{}, \qgsjet{} and \eposlhc{} event generators (shown as inset in \fig~\ref{fig:pp_pC_extrapolation}). The pp data points are multiplied with $R_p\text{C}$. The experimental and MC errors are geometrically summed (summing linearly marginally affects the resulting errors). The $p$C ``data points'' are then fitted with splines using the identical method as for the other cross sections in \ddm{}. 
  
    \section{Effective areas for phase space figures}
	 \label{app:effective_areas}
	 The neutrino telescope contours in \figs~\ref{fig:phase_space_neutrinos_muons} and \ref{fig:phase_space_prompt} are obtained from calculated event rates $R_\nu$ as a convolution of predicted neutrino fluxes with the effective areas of the experiments
	 \begin{equation}
	 	R_\nu(\Phi_\nu) = \int\rmd\cos{\theta}\int \rmd E_\nu~\Phi_\nu(E_\nu, \cos{\theta}) ~ \aeff(E_\nu, \cos{\theta}).
	 \end{equation}
	 The effective areas for IceCube detectors have been computed from histograms of Monte Carlo events from the IceCube Public Data Release \footnote{\url{https://icecube.wisc.edu/science/data-releases/}}. For the DeepCore and Upgrade $\aeff$'s, a cut on reconstructed energy of 60 GeV has been applied, in line with what has been used in oscillation analyses \cite{Abbasi:2011ym}.
For Super-Kamiokande a zenith-averaged approximate $\aeff$ has been obtained by reverse-engineering the spectra shown in Ref.~\cite{Abe:2017aap} assuming the flux from \mceq{} for \ddm{} + GSF.

  \section{Table of spectrum-weighted moments}
  \label{app:z-factor-table}
  \begin{table}
    \begin{tabularx}{0.8\textwidth}{YYYYY}
      $p$C, 31 GeV  & $\gamma_I = 1.0$ & $\gamma_I = 1.7$ & $\gamma_I = 2.0$ & $\gamma_I = 2.7$ \\
    \end{tabularx}
    \begin{tabularx}{0.8\textwidth}{XXXXX}
      \hline
      $\pi^+$ & 0.1634 $\pm$ 1.5\%  & 0.0477 $\pm$ 3.0\%  & 0.0306 $\pm$ 4.0\%  & 0.0126 $\pm$ 7.4\%  \\
      $\pi^-$ & 0.1120 $\pm$ 2.9\%  & 0.0305 $\pm$ 8.2\%  & 0.0193 $\pm$ 11.9\% & 0.0080 $\pm$ 23.8\% \\
      $K^+$   & 0.0194 $\pm$ 12.0\% & 0.0067 $\pm$ 25.0\% & 0.0047 $\pm$ 32.6\% & 0.0023 $\pm$ 53.0\% \\
      $K^-$   & 0.0055 $\pm$ 10.7\% & 0.0016 $\pm$ 26.3\% & 0.0010 $\pm$ 36.6\% & 0.0004 $\pm$ 70.2\% \\
      \hline
      \vspace{.25cm}
    \end{tabularx}
    
    \begin{tabularx}{0.8\textwidth}{YYYYY}
      $p$C, 158 GeV & $\gamma_I = 1.0$ & $\gamma_I = 1.7$ & $\gamma_I = 2.0$ & $\gamma_I = 2.7$ \\
    \end{tabularx}
    \begin{tabularx}{0.8\textwidth}{XXXXX}
      \hline
      $p$           & 0.2361 $\pm$ 3.0\%  & 0.1522 $\pm$ 4.0\%  & 0.1335 $\pm$ 4.4\%  & 0.1046 $\pm$ 5.3\%  \\
      $n$           & 0.1181 $\pm$ 11.6\% & 0.0747 $\pm$ 14.6\% & 0.0640 $\pm$ 16.1\% & 0.0477 $\pm$ 19.6\% \\
      $\pi^+$       & 0.1855 $\pm$ 7.3\%  & 0.0485 $\pm$ 16.8\% & 0.0310 $\pm$ 24.1\% & 0.0133 $\pm$ 47.8\% \\
      $\pi^-$       & 0.1310 $\pm$ 6.7\%  & 0.0267 $\pm$ 3.0\%  & 0.0154 $\pm$ 3.0\%  & 0.0052 $\pm$ 4.3\%  \\
      $K^+$ ($pp\to p$C) & 0.0188 $\pm$ 2.8\%  & 0.0050 $\pm$ 5.2\%  & 0.0031 $\pm$ 6.9\%  & 0.0012 $\pm$ 11.7\% \\
      $K^-$ ($pp\to p$C) & 0.0110 $\pm$ 4.9\%  & 0.0024 $\pm$ 3.5\%  & 0.0014 $\pm$ 4.0\%  & 0.0004 $\pm$ 7.4\%  \\
      $\bar{p}$ & 0.0043 $\pm$ 4.6\%  & 0.0010 $\pm$ 8.5\%  & 0.0005 $\pm$ 11.4\% & 0.0002 $\pm$ 21.4\% \\
      \hline
      \vspace{.25cm}
    \end{tabularx}
    
    \begin{tabularx}{0.8\textwidth}{YYYYY}
      $\pi^-$C, 158 GeV & $\gamma_I = 1.0$ & $\gamma_I = 1.7$ & $\gamma_I = 2.0$ & $\gamma_I = 2.7$ \\
    \end{tabularx}
    \begin{tabularx}{0.8\textwidth}{XXXXX}
      \hline
      $p$           & 0.0224 $\pm$ 1.3\%  & 0.0070 $\pm$ 2.9\%  & 0.0049 $\pm$ 3.9\%  & 0.0025 $\pm$ 6.3\%  \\
      $\pi^+$       & 0.1764 $\pm$ 1.5\%  & 0.0503 $\pm$ 0.7\%  & 0.0336 $\pm$ 0.9\%  & 0.0159 $\pm$ 1.7\%  \\
      $\pi^-$       & 0.3565 $\pm$ 13.9\% & 0.1556 $\pm$ 28.1\% & 0.1225 $\pm$ 34.0\% & 0.0810 $\pm$ 46.2\% \\
      $K^+$         & 0.0243 $\pm$ 0.8\%  & 0.0075 $\pm$ 1.9\%  & 0.0050 $\pm$ 2.6\%  & 0.0022 $\pm$ 4.7\%  \\
      $K^-$         & 0.0275 $\pm$ 0.9\%  & 0.0093 $\pm$ 1.8\%  & 0.0064 $\pm$ 2.3\%  & 0.0030 $\pm$ 3.6\%  \\
      $\bar{p}$ & 0.0168 $\pm$ 0.5\%  & 0.0066 $\pm$ 0.9\%  & 0.0048 $\pm$ 1.2\%  & 0.0025 $\pm$ 1.9\%  \\
      \hline
      \vspace{.25cm}
    \end{tabularx}
    \begin{tabularx}{0.8\textwidth}{YYYYY}
      $\pi^-$C, 350 GeV & $\gamma_I = 1.0$ & $\gamma_I = 1.7$ & $\gamma_I = 2.0$ & $\gamma_I = 2.7$ \\
    \end{tabularx}
    \begin{tabularx}{0.8\textwidth}{XXXXX}
      \hline
       $p$           & 0.0171 $\pm$ 3.7\%  & 0.0039 $\pm$ 3.9\%  & 0.0023 $\pm$ 4.9\%  & 0.0008 $\pm$ 7.5\%  \\
       $\pi^+$       & 0.1632 $\pm$ 2.4\%  & 0.0417 $\pm$ 6.0\%  & 0.0263 $\pm$ 7.9\%  & 0.0107 $\pm$ 13.0\% \\
       $\pi^-$       & 0.3394 $\pm$ 13.3\% & 0.1409 $\pm$ 27.7\% & 0.1086 $\pm$ 33.9\% & 0.0688 $\pm$ 47.0\% \\
       $K^+$         & 0.0183 $\pm$ $~~~$--  & 0.0039 $\pm$ $~~~$--  & 0.0022 $\pm$ $~~~$--  & 0.0006 $\pm$ $~~~$--  \\
       $K^-$         & 0.0220 $\pm$ $~~~$--  & 0.0056 $\pm$ $~~~$--  & 0.0034 $\pm$ $~~~$--  & 0.0012 $\pm$ $~~~$--  \\
       $\bar{p}$ & 0.0175 $\pm$ 12.5\% & 0.0061 $\pm$ 21.7\% & 0.0042 $\pm$ 26.0\% & 0.0ı019 $\pm$ 36.3\% \\
      \hline
    \end{tabularx}
    \caption{Table of spectrum-weighted moments computed for $p$C from data of the NA61 experiment at 31 GeV, NA49 $pp$ and $p$C at 158 GeV, and NA61 for $\pi^-$C at 158 and 350 GeV beam momenta as listed in Table~\ref{tab:na_data}. For $K^\pm$ secondaries in $\pi^-$C collisions at 350 GeV the error on the spectrum weighted integral is unconstrained.\label{tab:$z$-factors}}
  \end{table}
Numerical computed spectrum-weighted moments or $Z$ factors are shown in Table~\ref{tab:$z$-factors}. The integral spectral index is related to the that of the cosmic-ray nucleon flux as $\gamma_I = \gamma - 1$. More tabulated $Z$ factors can be found in Table~5.2 of Ref.~\cite{Gaisser:2016cr}.

\end{appendix}

\end{document}